\documentclass[manuscript,11pt]{aastex}
\usepackage{epsfig}

\usepackage{soul}

\def\oviii{O\,{\sc viii]}}
\def\ovii{O\,{\sc vii]}}

\def\fexvii{Fe\,{\sc xvii}}

\def\fexxv{Fe\,{\sc xxv}}
\def\fexxvi{Fe\,{\sc xxvi}}

\def\arxviii{Ar\,{\sc xviii}}
\def\arxvii{Ar\,{\sc xvii}}

\def\sxvi{S\,{\sc xvi}}
\def\sxv{S\,{\sc xv}}
\def\sixiv{Si\,{\sc xiv}}
\def\sixiii{Si\,{\sc xiii}}

\def\mgxii{Mg\,{\sc xii}}
\def\nex{Ne\,{\sc x}}
\def\alxiii{Al\,{\sc xiii}}

\def\civ{C\,{\sc iv}}

\def\ovii{O\,{\sc vii}}
\def\oviii{O\,{\sc viii}}

\def\mgxii{Mg\,{\sc xii}}
\def\civ{C\,{\sc iv}}
\def\nvii{N\,{\sc vii}}
\def\ovii{O\,{\sc vii}}
\def\nv{N\,{\sc v}}

\def\cm{\ifmmode {\rm cm}^{-1} \else cm$^{-1}$ \fi}
\def\s{\ifmmode {\rm s}^{-1} \else s$^{-1}$ \fi}
\def\cc{\ifmmode {\rm cm}^{-3} \else cm$^{-3}$ \fi}
\def\cs{\ifmmode {\rm cm}^{-2} \else cm$^{-2}$ \fi}
\def\g{\ifmmode \gamma \else $\gamma$\fi}
\def\G{\ifmmode \Gamma \else $\Gamma$\fi}
\def\Gs{\ifmmode \Gamma~ \else $\Gamma~$\fi}

\def\gc{\ifmmode \gamma_{\rm c} \else $\gamma_{\rm c}$ \fi}
\def\sw{Schwarzschild~}
\def\gsim{\mathrel{\raise.5ex\hbox{$>$}\mkern-14mu
             \lower0.6ex\hbox{$\sim$}}}
\def\lsim{\mathrel{\raise.3ex\hbox{$<$}\mkern-14mu
             \lower0.6ex\hbox{$\sim$}}}
\def\simless{\mathbin{\lower 3pt\hbox
     {$\rlap{\raise 5pt\hbox{$\char'074$}}\mathchar"7218$}}}   %< or of order
\def\simmore{\mathbin{\lower 3pt\hbox
     {$\rlap{\raise 5pt\hbox{$\char'076$}}\mathchar"7218$}}}   %> or of order
\def\Msun{M_\odot}                                % solar masses
\def\4u{4U 1728--34}
\def\deg{^\circ}
\def\aa{\buildrel _{\circ} \over {\mathrm{A}}}

\def\obj{NGC~3783}

\lefthead{et al.} \righthead{\4u}

\shorttitle{MHD-Driven WAs in NGC~3783}

\shortauthors{Fukumura et al.}

\begin{document}

%\title{A Magnetic View of Warm Absorbers in NGC~3783}
\title{Magnetized Disk-Winds in NGC~3783}
%\title{Soft X-Ray Excess from General Relativistic Magnetohydrodynamic Accreting Plasma}
%\title{Comptonized Spectra from Relativistic Accreting Plasma}

\date{\today}

\author{\textsc{Keigo Fukumura}\altaffilmark{1,2},
\textsc{Demosthenes Kazanas}\altaffilmark{3}, \textsc{Chris Shrader}\altaffilmark{3,4},
\textsc{Ehud Behar}\altaffilmark{5}, \textsc{Francesco Tombesi}\altaffilmark{3,6,7}
\textsc{and} \textsc{Ioannis Contopoulos}\altaffilmark{8} }

\altaffiltext{1}{Department of Physics and Astronomy, James Madison University,
Harrisonburg, VA 22807; fukumukx@jmu.edu}
%\altaffiltext{2}{Email: fukumukx@jmu.edu}
\altaffiltext{2}{KITP Scholar at UC Santa Barbara }
\altaffiltext{3}{Astrophysics Science Division, NASA/Goddard Space Flight Center,
Greenbelt, MD 20771}
%\altaffiltext{5}{Universities Space Research Association, 7178 Columbia Gateway Dr. Columbia, MD %21046}
\altaffiltext{4} {Catholic University of America, Washington, DC 20064}
\altaffiltext{5}{Department of Physics, Technion, Haifa 32000, Israel}
\altaffiltext{6}{Department of Astronomy and CRESST, University of Maryland, College
Park, MD20742}
\altaffiltext{7}{Department of Physics, University of Rome ``Tor
Vergata", Via della Ricerca Scientifica 1, I-00133 Rome, Italy}
\altaffiltext{8}{Research Center for Astronomy, Academy of Athens, Athens 11527,
Greece}

\begin{abstract}

\baselineskip=15pt

We analyze a 900-ks stacked {\it Chandra}/HETG spectrum of \obj\ in the context of
magnetically-driven accretion-disk wind models in an effort to provide tight
constraints on the global conditions of the underlying absorbers. Motivated by the earlier measurements of its absorption measure distribution (AMD) indicating X-ray-absorbing ionic columns that decrease
slowly with decreasing ionization parameter, we
employ 2D magnetohydrodynamic (MHD) disk-wind models to describe the
global outflow. We compute its photoionization structure along with the wind
kinematic properties allowing us to further calculate in a self-consistent
fashion the shapes of the major X-ray absorption lines. With
the wind radial density profile determined by the AMD, the profiles of the ensemble
of the observed absorption features are determined by the two global parameters of the
MHD wind; i.e. disk inclination $\theta_{\rm obs}$ and wind density normalization $n_o$.
Considering the most significant absorption features in the ($\sim 1.8\aa - 20\aa$) range, we
show that the MHD-wind is best described by $n(r) \sim
6.9 \times 10^{11} (r/r_o)^{-1.15}$ cm$^{-3}$ and $\theta_{\rm obs}=44 \deg$. We
argue that winds launched by X-ray heating, radiation pressure or even MHD
winds but with steeper radial density profiles are strongly disfavored by data.
Considering the properties of Fe K band absorption features (i.e. \fexxv\ and \fexxvi), while typically prominent in the AGN X-ray spectra, they appear to be weak in \obj. 
For the specific parameters of our model obtained by fitting the AMD and the rest of absorption features, these features are found to be weak in agreement with observation.

\end{abstract}


%\keywords{accretion, accretion disks --- galaxies: Seyfert ---
%shock waves --- (magnetohydrodynamics:) MHD  --- X-rays:
%galaxies}

\baselineskip=15pt

\section{Introduction}

Ionized outflows are a common feature of Active Galactic Nuclei (AGN), manifesting
themselves as blue-shifted absorption spectral lines. Approximately $50\%-70\%$ of all
Seyfert 1's exhibit such features in their UV spectra and a similar fraction in their
X-ray spectra \citep{CKG03}. The most likely process responsible for the outflowing
plasma ionization is photoionization by the AGN ionizing continuum; in this respect,
X-rays appear to be of broader utility in probing their properties since X-ray
transitions span a much larger range in photoionization parameter $\xi$ than the UV
ones.

X-ray absorption features were discovered first in the {\em Einstein} QSO spectra
\citep[e.g.][]{Halpern84}, with more significant detections of K-shell absorption edges
due to \ovii\  (0.74 keV) and \oviii\ (0.87 keV)  by {\it ROSAT} \citep{NandraPounds92, Nandra93, Fiore93, Turner93}. Later on, the
improved spectroscopic capabilities of {\it ASCA} confirmed the robust presence of
these edges in many luminous Seyfert AGNs \citep[e.g.][]{Reynolds97,George98};
attributed to absorption by warm plasmas ($T \sim 10^6$ K), they have been thenceforth
referred to as Warm Absorbers (WA). With the much enhanced spectral resolution and
senstitivity of dispersive spectrometers onboard {\it Chandra} and {\it XMM-Newton}, it
became obvious that there exists a plethora of absorption lines of various
charge states of many elements; these span a wide range of ionization parameter $\xi
\equiv L_{\rm ion}/(n r^2)$ \citep[e.g.][]{CKG03,Blustin05,Steenbrugge05,McKernan07} where $L_{\rm ion}$ is the
ionizing (X-ray) luminosity, $n$ is electron number density at radius $r$. Their
columns lie in the range $10^{20} \lesssim N_H \lesssim 10^{22}$ cm$^{-2}$, their
ionization parameter between $-1 \lesssim \log \xi \lesssim 4$, their temperatures
between $10^4 < T  <10^7$ and the exhibit line-of-sight (LoS) velocities of $v/c \lesssim 0.01$
\citep[e.g.][]{ReynoldsFabian95}, implying distances $r \gsim 10^4$ \sw radii,
employing the Keplerian association between velocity and radius.

\obj\ is a nearby ($z=0.00976$),  bright AGN  with black hole mass of $3 \times 10^7
\Msun$ \citep[][]{Peterson04}; it has been observed with {\it Chandra}/HETGS a number
of times to date since 2000 \citep[][]{Kaspi00}, and also in conjunction with
simultaneous observations by {\it ASCA} and {\it RXTE} \citep[][hereafter K01]{K01}.
It is one of the most intensively  monitored Seyfert galaxies for its high-resolution
absorption study with a total duration of $900$ ks {\it Chandra} grating data
\citep[i.e. five $\sim 170$ ks observations and a $56$ ks one;][hereafter, K02]{K02}.
The mean X-ray luminosity in 2-10 keV is $L_X = 3 \times 10^{43}$ erg~s$^{-1}$ (K01 and K02).
In addition to a series of ionized absorbers in X-ray, \obj\ is also known to exhibit
UV absorbers detected with {\it FUSE} and  {\it HST}/STIS \citep[e.g.][and references
therein]{KraemerCrenshawGabel01}.  In particular, K02 conducted an exploratory spectral
analysis of the detected X-ray absorbers identifying the physical characteristics of
individual ions in the broad-band stacked spectrum assuming phenomenological multiple
absorption systems. They did not find any correlation of their velocity shifts or their
FWHMs with ionization potentials. Along similar lines, \citet{Krongold03}
\citep[also][]{Netzer03} analyzed the same 900 ks {\it Chandra} grating spectrum with a
self-consistent photoionization model, assuming a simple geometry that consists of a
central source emitting an ionizing SED and clouds of gas intercepting our line of
sight. Employing {\tt cloudy} \citep[version 90.04;][]{Ferland13} to obtain the
individual clouds' ionization state, they were able to constrain the physical
parameters of various ions, i.e. their ionization parameter $\xi$, column $N_H$,
outflow velocity $v$ and internal turbulent velocity $v_{\rm turb}$.

In addition to these {\em Chandra} observations, \obj\ was observed with  {\it
XMM-Newton}/EPIC over two complete satellite orbits in 2001 (ID 0112210501 and
0112210201) with a total good exposure of 280 ks with gratings \citep{Behar03} and
$\sim 240$ ks with CCDs to study the Fe K line profile \citep[][]{Reeves04}. As a part
of their spectral analysis, it was found that the highest ionization component present
during this epoch is at an ionization of $\log \xi \simeq 3$ and column density of $N_H
\sim 5 \times 10^{22}$ cm$^{-2}$ at an estimated  distance of $r < 0.1$ pc from the
nucleus, while the low ionization states are many pc away. \cite{Gabel05} confirmed the
large distances of the absorbers (with low densities) by observations of $C^{+2}$ transitions from excited
levels.
More recently, \cite{Mehdipour17} conducted a {\it Swift} monitoring campaign
triggering joint observations with {\it XMM-Newton}, {\it NuSTAR} and {\it
HST}/COS/STIS, in cases of hard \emph{Swift}-XRT spectra, indicating soft X-ray
absorption. It was found, in contrast to the previous analyses performed in the
2000-2001 (unobscured) epoch, that the central X-ray region was heavily obscured by
outflowing plasma which, besides absorbing the low energy ($E \lsim 1$ keV) X-rays,
exhibited also deep absorption features very distinct from the absorbers detected in
the previous UV/X-ray observations. In particular, high-velocity \fexxv\ and \fexxvi\
absorbers ($v_{\rm out} \sim$ a few thousand km~s$^{-1}$) of high column density ($N_H \sim
10^{23}$ cm$^{-2}$) are clearly present.
%1
While comprehensive in their analysis  methodology, the physical realization of the
series of absorbers observed was still unclear consisting of multiple absorber
components as has been often invoked, and the global identification of the observed
outflows of many ions is not explicitly revealed.
%
%Toward a Self-Consistent Model of the Ionized Absorber in NGC 3783
%Authors:	
%Krongold, Y.; Nicastro, F.; Brickhouse, N. S.; Elvis, M.; Liedahl, D. A.; Mathur, S.

The presence of ionic species in absorption in the AGN spectra with a broad range in
ionization parameter $\xi$ has been generally dealt by considering a number of separate
components of well defined $\xi$. An altogether different (and profitable) approach has
been that of \cite{HBK07}; these authors, noticing the broad range in $\xi$ of the
ionic species in the data, assumed a continuous distribution of hydrogen
equivalent ionic columns $N_H$ on $\xi$ of the form $N_H \propto \xi^{\alpha}$ (or more
precisely the distribution of $d N_H/ d\log \xi$, their so-called Absorption Measure
Distribution or AMD as discussed below in equations~(\ref{eq:amd1})-(\ref{eq:amd2})). Then, through a minimization procedure, they were able to consolidate
the ensemble of the properties of all transitions into the value of single parameter,
namely $\alpha$, which, surprisingly, was found to have a very limited range $\alpha
\simeq 0.01 - 0.3$ in the number of AGN with data of sufficiently high quality for such
an analysis \citep[][hereafter B09]{B09}. This behavior has been since found in a joint
analysis of twenty six Seyfert outflows (Laha et al. 2014).

While the largest possible location $R_{\rm max}$ of an absorber can be estimated by $R_{\rm max} = L_{\rm ion}/(\xi N_H)$, by recasting the ionization parameter in a slightly different form based on the LoS-integrated hydrogen number density, one can derive a simple analytic expression for a local finite column density $\Delta N_H$ for a finite ionization parameter bin $\Delta \xi$ over a small radial LoS extent $\Delta r$ as
\begin{eqnarray}
\Delta N_H = n(r) \Delta r \propto \xi^{(3-2p)/(p-2)} \Delta \xi \ , \label{eq:amd1}
\end{eqnarray}
where $n(r) \propto r^{-p}$ is the global wind density profile along a LoS. One can then derive the expected AMD as a function of $\xi$ as
\begin{eqnarray}
\textmd{AMD} \equiv \lim_{\Delta r \to 0} \frac{\Delta N_H}{\Delta \left(\log \xi \right)} \propto \xi^{-(p-1)/(p-2)}  \ , \label{eq:amd2}
\end{eqnarray}
%
%i.e. $\xi = L/n(r)
%r^2 = L/N_H(r) r$,
as similarly derived in B09 and \cite{K12}.
One should note that, for a continuous distribution of $N_H$ on
$\xi$, measurement of $N_H$ of an ion of known $\xi$ provides a measure of its distance
$r$ from the ionizing source; repeating this process for ions of a wide range of $\xi$
can then provide the distribution of plasma density $n(r)$ along the observer's LoS.
Thus,
%the relation $N_H \propto \xi^{\alpha}$ implies $n(r) \propto r^{-p}$
%with $\alpha = (p-1)/(2-p)$ or $p = (2 \alpha +1)/(\alpha +1)$, so that
the observed slope of AMD implies plasma radial density profiles with a rather limited
range in their slopes, namely $1.02 \lesssim p \lesssim 1.22$ (see B09 for 5 Seyfert 1 AGNs including \obj).

Motivated by these considerations we have in the past modeled the AMD observations
within the framework of the 2D MHD winds of \citet[][hereafter, CL94]{CL94}, generalizations of those of
\citet[][hereafter, BP82]{BP82}, that allow a wider range of wind density profile along the LoS. We found
that detailed treatment of the photoionization of MHD winds with $p \simeq 1$ presented
a good approximation to the observed AMD dependence and velocity properties of the
Seyfert 1's in the list of B09 \citep[][hereafter, F10a]{F10a}; we also found that reduction of the
ionizing  X-ray content in the AGN SED, as is appropriate with the BAL QSO spectra,
provided velocities consistent with those observed in this AGN class \citep{F10b}.
Furthermore, considering that our 2D MHD winds are scale invariant \citep{K12}, we have
applied the same models to the high S/N {\it Chandra} spectra of the galactic XRB, GRO~J1655-40
\citep{F17}, to show that our models provide excellent fits to the absorption features
(both in absorption depth and in velocity) of the scaled down wind of this stellar
black hole.
In the context of a mutual interaction between accretion and ejection physics, for example, other groups have also investigated a physical constraint on a global structure of MHD-driven outflows \citep[e.g.][]{Ferreira97,Casse00a, Casse00b,Chakravorty16}.

In the present work, we employ the same 2D MHD wind model as in our previous works to
analyze the 900 ks {\em Chandra} HETG data of \obj, a well studied, nearby, radio-quiet
AGN. With its AMD already determined in B09, our emphasis is the precise
determination of the large scale wind parameters, namely the value of the index $p$,
its inclination angle and the wind density normalization. This we do by providing
detailed fits to its most significant absorption lines. In \S 2 we provide a brief
outline of the MHD winds and a comparison with other outflow models. In \S 3 we
describe our analysis procedure. Our results and their comparison to observation are
shown in \S 4 demonstrating that the wind model can describe the observations
successfully. We summarize and discuss the implications of the model in \S 5.

%In this work, we analyse the \obj\ 900 ks data by employing the same wind model   to
%\obj\ (one of the sources in B09), a well-studied radio-quiet nearby Seyfert, in an
%attempt to understand the observed WAs from a broader perspective hinted by the derived
%flat AMD. The goal of the present study is to self-consistently model the observed
%ionized absorbers in \obj\ within a unique description of magnetic winds based in a
%simplistic approach with a fewer parameters. We aim to better describe the absorbers
%both at macroscopic level (i.e. AMD and wind correlations) as well as microscopic level
%(i.e. individual spectral fitting for each ion) and also constrain its large-scale wind
%structure. In this paper, we briefly the observations of \obj\ in \S 1, followed
%by describing the generic feature of the MHD-driven accretion-disk winds in \S 2. In \S
%3 we show our analysis and results by successfully demonstrating that the wind model
%can account for observations. We summarize and discuss the implications of the model in
%\S 4.

\section{Overview of MHD-Driven Wind Model}

Given the large bolometric luminosity of AGN and the X-ray contribution to their
continua,
%Given the large bolometric luminosity of AGN and the X-ray content their
%continua,
it is natural to consider radiation pressure \citep{MCGV95,PSK00} and/or
X-ray heating \citep{BMS83} as the agents that drive their ubiquitous outflows. While
one cannot give a preference to MHD launching over these processes \emph{a priori}, the
general argument in favor of the latter and against the simplest versions of the former
comes from the form of AMD: For both these processes (i.e. radiation pressure and X-ray
heating), at some distance a few times larger than the size of their driving source,
these winds will look quasi-spherical, with their velocity increasing with $r$ to its
asymptotic value. Then, in their acceleration region $(dv/dr>0)$, due to mass
conservation, their densities should decrease faster than $r^{-2}$ resulting in $N_H$
decreasing with increasing $\xi$ \citep{Luketic10}. Such a behavior is contrary to that
observed in the compilation of B09 and \cite{Laha14}. Perhaps more involved models
could reproduce the observed $N_H - \xi$ behavior, however, we are not aware of any so
far.

On the other hand, the broad range of $\xi$ observed in the data suggests a
self-similar process that spans several decades in $\xi$ and $r$. The 2D winds of BP82 and CL94 serve as a reasonable guess to this end, as they are
launched over the entire disk extent. Their velocities (radial and azimuthal) scale
with the Keplerian one, $v_{\rm out} \propto r^{-1/2}$; their densities are separable
in $r$ and $\theta$ (due to self-similarity) and take the form $n(r,\theta) = n_{o}
(r/r_S)^{-p} f(\theta)$; $n_o$ denotes the density normalization (i.e. wind density at
its innermost launching radius on the disk surface at $r \gsim r_S;~r_S$ is the black
hole \sw radius) and it can be expressed in units of dimensionless mass flux rate $\dot
m = \dot M/\dot M_{\rm Edd}$ by
\begin{eqnarray}
n_o \sim \dot m/(\sigma_T r_S) \ ,
\end{eqnarray}
where we assume that accreting mass is equally distributed between accretion and outflows at each radius (i.e. $f_w=1$ as in F10a; \citealt{F17}).
The function $f(\theta)$ determines the angular dependence of the wind and it is given
by the solution of the Grad-Shafranov equation (see CL94). It has a steep
$\theta$-dependence [an approximate expression is $f(\theta) \sim e^{5(\theta -
\pi/2)}$, see Fig.~2 of F10a], giving the winds a toroidal appearance. Because
of this feature it was suggested \citep{KK94} that such winds are in fact the AGN tori
invoked to account for AGN unification. The precise form of $f(\theta)$, i.e. the
winds' opening angle, depends on their specific angular momentum \citep{F14}. However,
it is qualitatively similar to the form given above.

More importantly, the density radial dependence index $p$ above is  intimately related
to the AMD shape by equation~(\ref{eq:amd2}).
%%
%\begin{eqnarray}
%N_H \propto \xi^{-\left(p-1 \right)/\left(p-2 \right)} \ .
%\end{eqnarray}
%%
The wind velocity, mainly in the $\phi$-direction as $\theta \rightarrow 90 \deg$,
becomes mainly radial at larger latitudes (see Fig.~1a in F10a). Its projection
along the observer's LoS depends on the disk inclination and affects the shape of
absorption features; however, due to their Keplerian scaling the outflow velocity
component scales with ionization parameter $\xi$ like
\begin{eqnarray}
v_{\rm out} \propto \xi^{1/\{2(2-p)\}} \ , \label{eq:vel}
\end{eqnarray}
a feature that figures prominently in the shapes of the absorption line profiles (see
F10a and \S 4.1).

The winds have the following generic features: (1) The wind structure (i.e. the
dependence of $N_H$ on $\theta$ and on $r/r_S$) is fundamentally mass invariant, so
they are applicable across the black hole mass scales. (2) They extend from near the
black hole ISCO ($r \gsim 3 \, r_S$) to the outer disk edge, with velocities that
decrease like $r^{-1/\,2}$. The first feature is indicative of the universal presence
of magnetically-launched ionized winds in both AGNs (e.g., \citealt{Couto16}; \citealt{KraemerTombesi17}  for NGC~4151 and \citealt{Turner05} for NGC~3516) as well as XRBs as often claimed
\citep[e.g.][]{Miller06,Miller08, Kallman09,Miller15,F17}; the second feature clearly
points to a strong relevance to the so called ultra-fast outflows (UFOs), increasingly
discovered in many Seyfert AGNs \citep[e.g.][]{Reeves09,Tombesi10a,Tombesi13,Tombesi15,Gupta15,Gofford15} and
gravitationally lensed quasars \citep[e.g.][]{Chartas09a}, and their relation to the
lower velocity WAs. 
Especially, \cite{KraemerTombesi17} has made a strong argument to support MHD-driven scenario to explain the UFOs detected in NGC~4151 in an approach similar to ours. 
As noted above, while the wind is present at all radii and their
corresponding velocities, its full ionization in the black hole vicinity leaves no
absorption imprints in the spectra. These first occur when the wind ionization drops
sufficiently to allow the presence of \fexxvi, \fexxv; the radii at which these ions
first occur depend on the contribution of the ionizing X-rays in the object's SED. The
radii are small in X-ray weak (BAL QSOs) and large in X-ray strong (galactic XRBs)
objects, yielding correspondingly large and small velocities for these ions.

The wind photoionization is computed as detailed in F10a, by employing {\tt
xstar} \citep[][]{KallmanBausas01}  to determine the plasma ionization and the relevant
cross sections to be used in the radiation transfer. Finally, we compute absorption
line profiles by calculating photo-excitation cross  section
\begin{eqnarray}
\sigma_{\rm abs} = 0.01495 (f_{ij} / \Delta \nu_D) H(a,u) \ , \label{eq:sigma}
\end{eqnarray}
as a function of local wind velocity  $v(r,\theta)$ and its radial shear $\Delta v_{\rm
sh}(r,\theta)$ through the Voigt function $H(a,u)$,
%given by
%%
%\begin{eqnarray}
%H(a,u) \equiv \frac{a}{\pi} \int_{-\infty}^\infty \frac{e^{-y^2}
%dy}{(u-y)^2+a^2}   \ ,   \label{eq:voigt}
%\end{eqnarray}
%%
where $f_{ij}$ is the oscillator strength  of the transition between the $i-$th and
$j-$th levels of an ionic species and $\Delta \nu_D$ is the Doppler broadening factor
estimated by $\Delta \nu_D \approx (\Delta v_{\rm sh}/c) \nu_0$ relative to the
centroid (rest-frame) frequency $\nu_0$.
With $\sigma_{\rm abs}$ in equation~(\ref{eq:sigma}), one can calculate the line depth $\tau$ as
\begin{eqnarray}
\tau = \sigma_{\rm abs} N_{\rm ion} \ , \label{eq:tau}
\end{eqnarray}
where $N_{\rm ion}$ is ionic column density computed with {\tt xstar} calculations \citep[see, e.g.][]{F15}.
In this formalism, therefore, the line
broadening is provided by the natural shear of the wind velocity field and, as such, we
eschew the use of the turbulent velocity parameter {\tt vturb}; most importantly, all
line profiles are computed using the local values of \emph{ the same} global wind
parameters rather than being treated as mutually independent multiple gaussian
functions, thereby over-constraining our models.

\begin{figure}[ht]% ------------------------------------- Figure~1
\begin{center}$
\begin{array}{cc}
%%\includegraphics[trim=0in 0in 0in
%&0in,keepaspectratio=false,width=3.2in,angle=-0,clip=false]{NGC3783_HETG_900ks.pdf}
\includegraphics[trim=0in 0in 0in
0in,keepaspectratio=false,width=4in,angle=-0,clip=false]{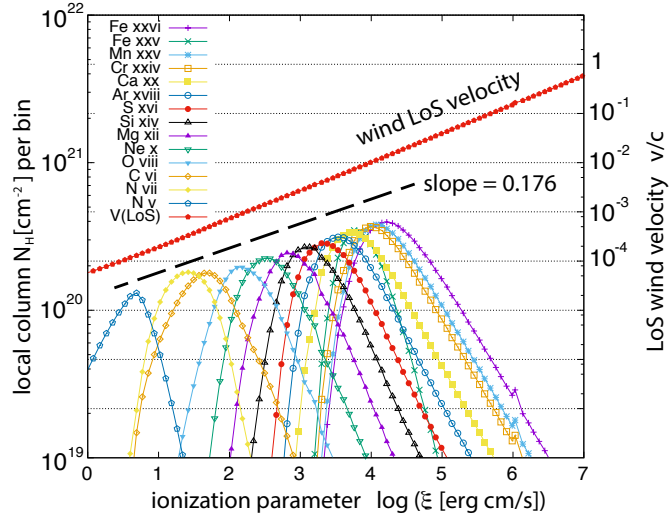}
\end{array}$
\end{center}
\caption{%(a) A stacked 900 ks {\it Chandra}/HETG/MEG spectrum of \obj\ studied in \cite{K02}.
An example of synthetic AMD of various ions modeled with the MHD-wind
of density profile of $p=1.15$ for a fiducial set of wind parameters with $\theta=40\deg$ and $n_{11}=13.6$. } \label{fig:amd}
\end{figure}

%\section{Methods}

\section{Analysis}

For the present study, we use the 900-ks stacked spectrum  from K02 (see their Fig.~5).
Our analysis is primarily focused on the $\sim 1\aa$ (Fe) $- 20\aa$ (O) broad-band
spectral fitting with the Galactic absorption of $N_H^{\rm Gal} = 8.7 \times 10^{20}$
cm$^{-2}$ \citep[e.g.][]{Alloin95}. Closely following K01 and K02, we estimate the underlying continuum
flux in discrete energy bands where no or little line signatures are present (i.e.
line-free-zones or LFZs). The continuum spectral shape is assumed to be locally linear
instead of a polynomial form, which seems to be quite acceptable, as also claimed in
K02, our aim in the present work not being a physical model of the AGN continuum.
Contrary to most works which assume a mutually-decoupled, discrete number of ionization and kinematic
components to fit groups of lines of similar values of $\xi$, our models provide a
continuum of ionization parameter, column and velocity projection along the LoS with
$r$, the distance from the AGN. As a result, our models are far more constrained
with no guarantee that they will provide the correct column (EW) and velocity at the
proper values of $\xi$ that support specific ions.
%
%With the estimate of the continuum  flux across a given absorption feature for many
%ions, our wind model will imprint absorption lines systematically for all the ions
%without the need to assume a discrete number of ionization zones and/or kinematic
%components typically adopted in phenomenological approach.
%%
%As a result, we are aware of a potential  uncertainty in modeling the absorption
%features, including its equivalent width (EW), due to this approach.

%
\begin{figure}[ht]% ------------------------------------- Figure~2
\begin{center}$
\begin{array}{ccc}
%\includegraphics[trim=0in 0in 0in
%0in,keepaspectratio=false,width=4.0in,angle=-0,clip=false]{bestfit_r115_el.pdf}
\includegraphics[trim=0in 0in 0in
0in,keepaspectratio=false,width=2.2in,angle=-0,clip=false]{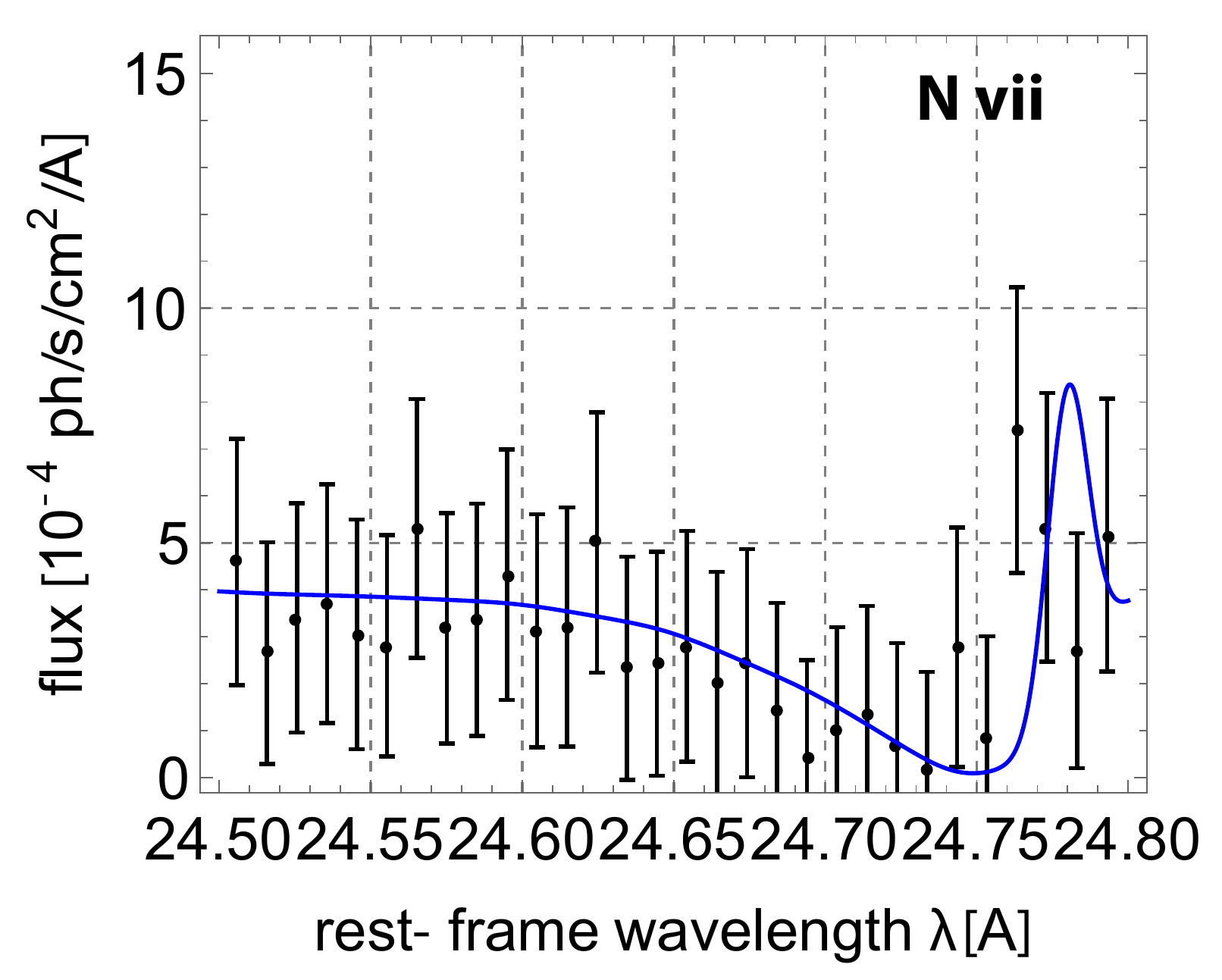}
\includegraphics[trim=0in 0in 0in
0in,keepaspectratio=false,width=2.2in,angle=-0,clip=false]{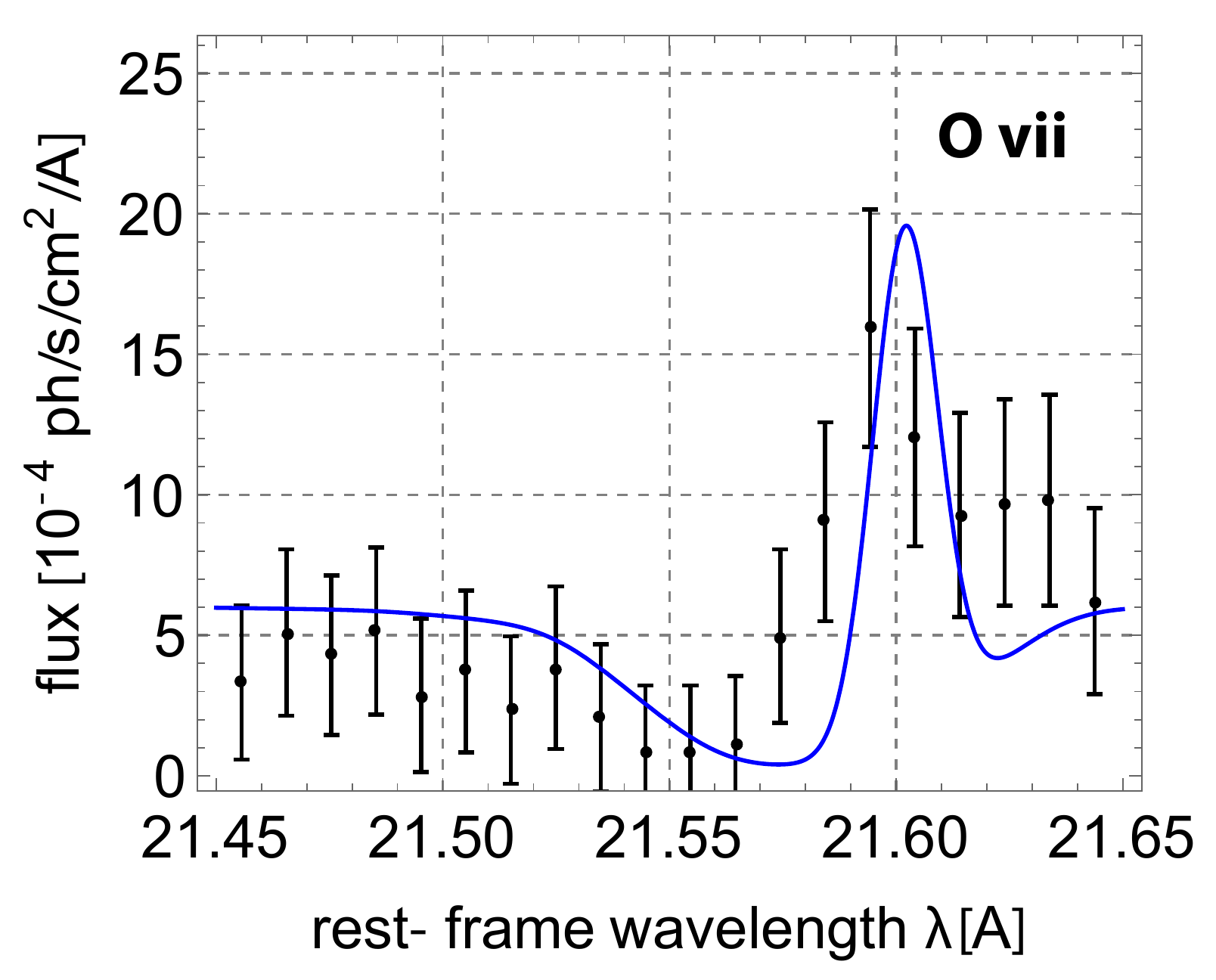}
\includegraphics[trim=0in 0in 0in
0in,keepaspectratio=false,width=2.2in,angle=-0,clip=false]{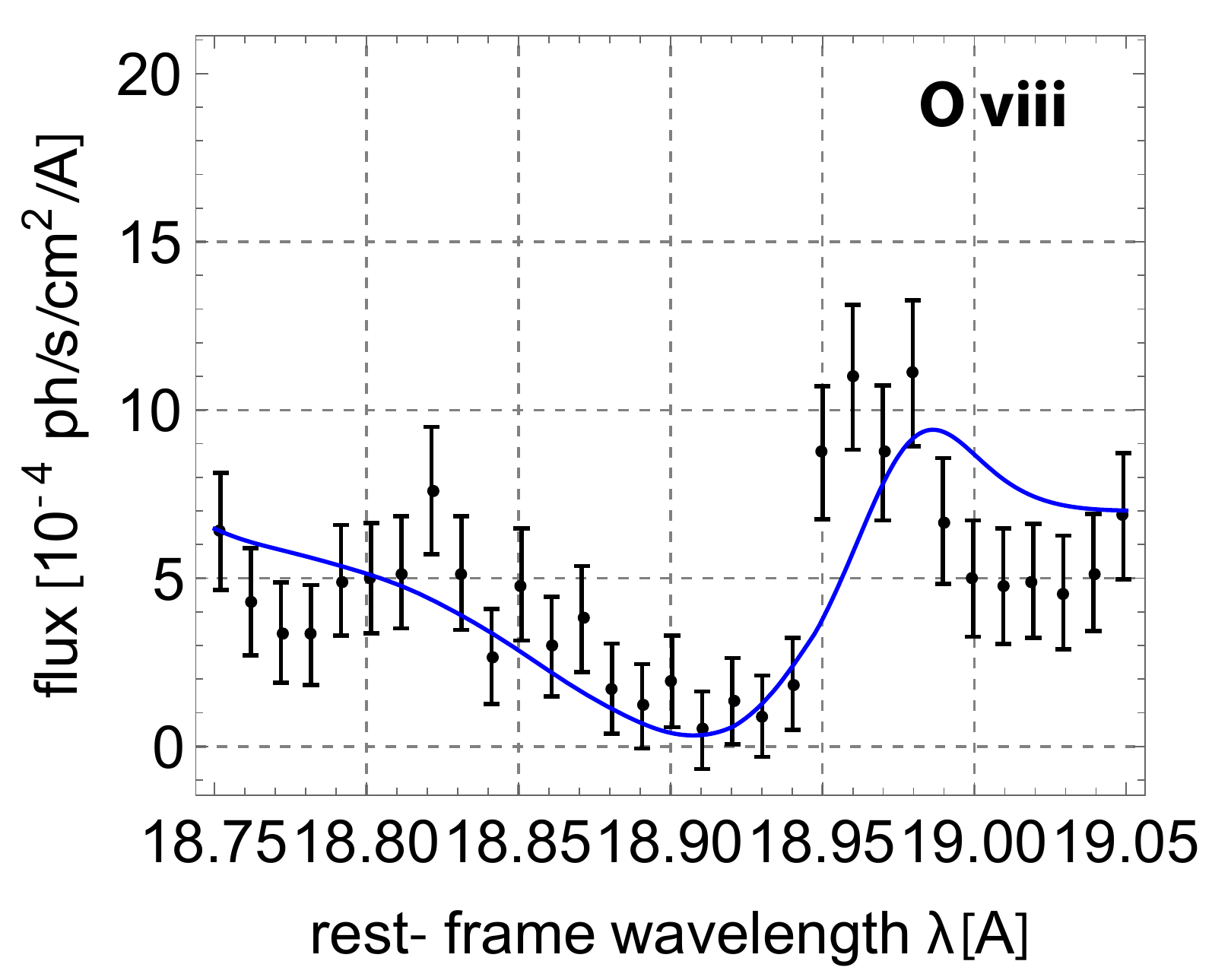}
\\
\includegraphics[trim=0in 0in 0in
0in,keepaspectratio=false,width=2.2in,angle=-0,clip=false]{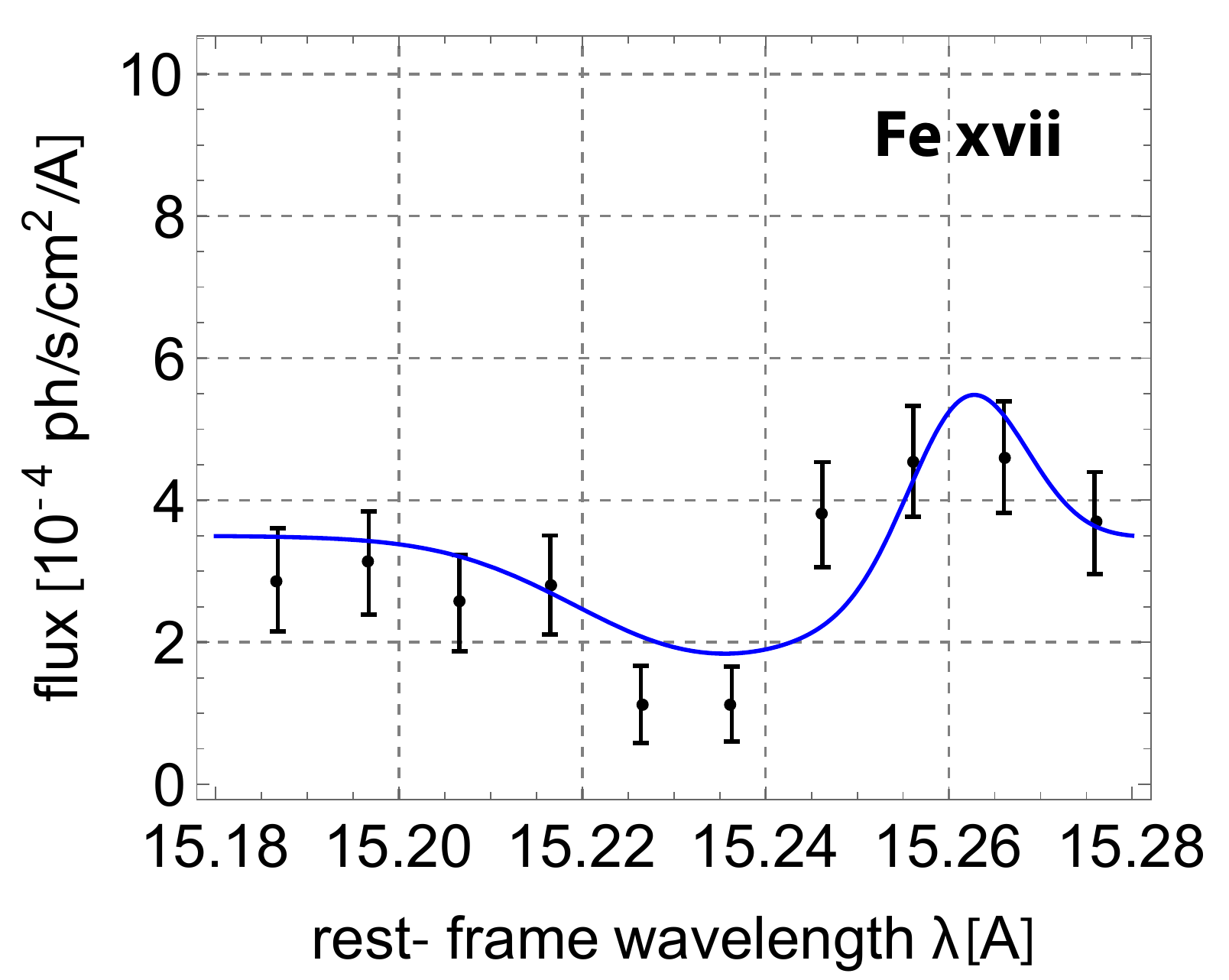}
\includegraphics[trim=0in 0in 0in
0in,keepaspectratio=false,width=2.2in,angle=-0,clip=false]{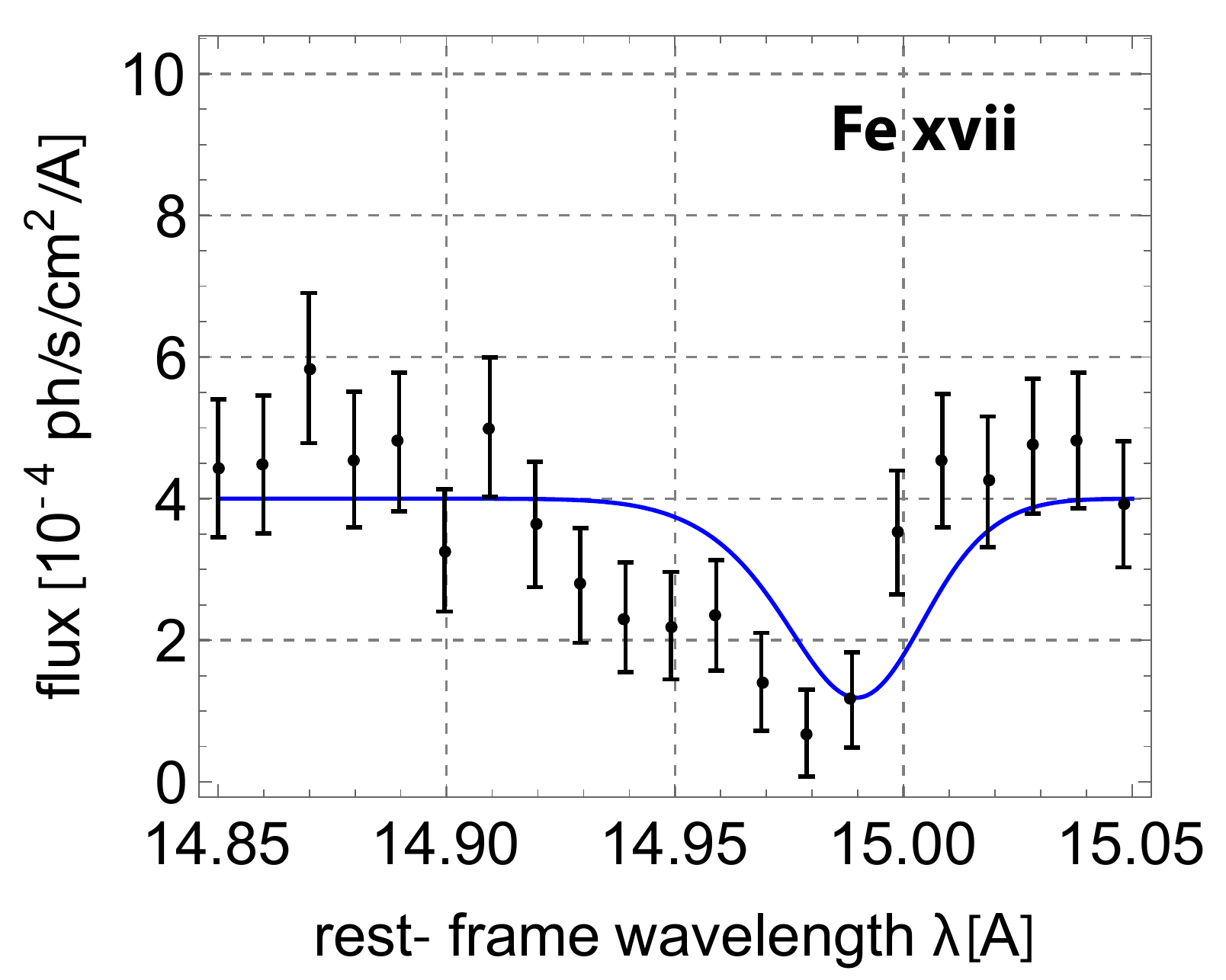}
\includegraphics[trim=0in 0in 0in
0in,keepaspectratio=false,width=2.2in,angle=-0,clip=false]{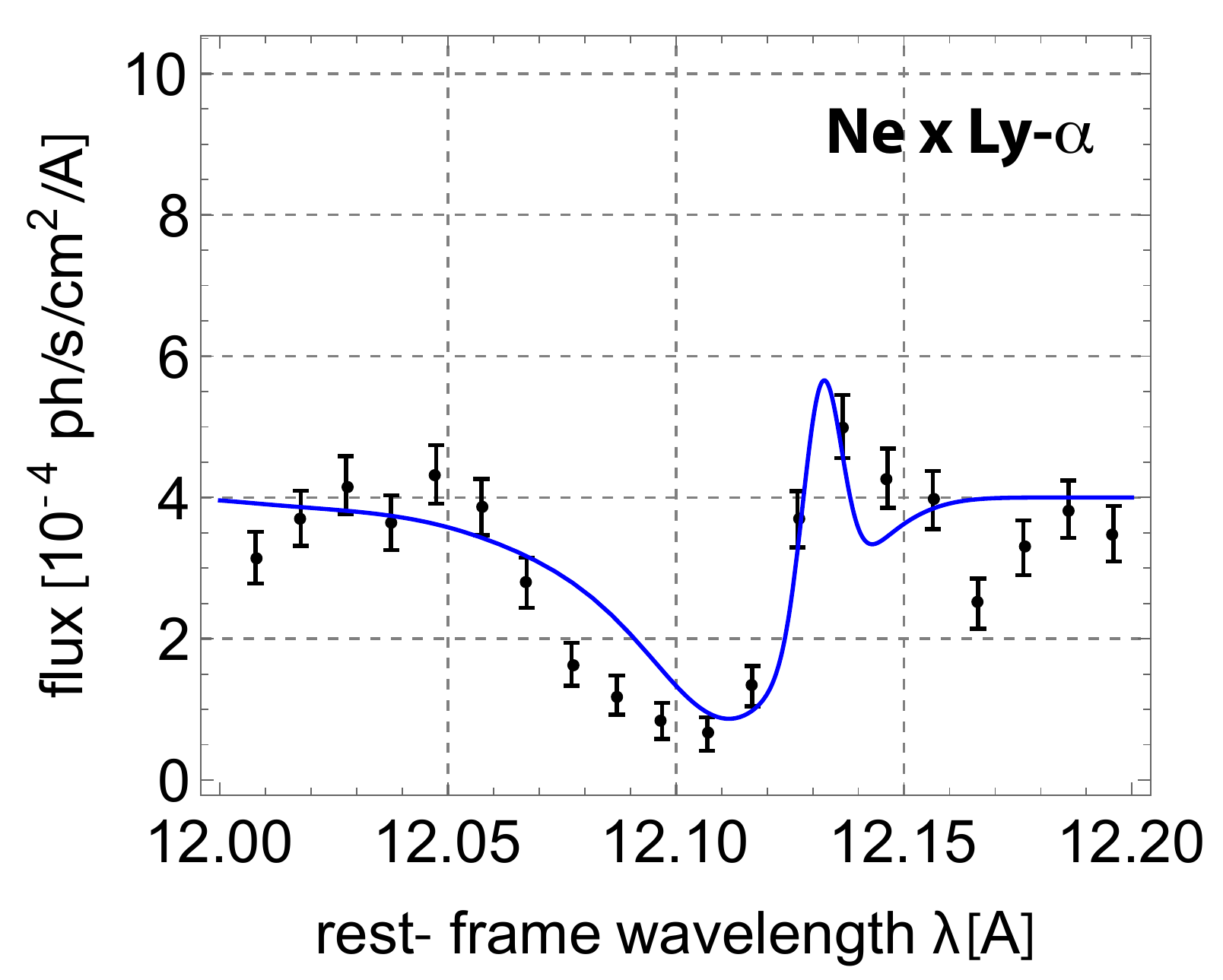}
\\
\includegraphics[trim=0in 0in 0in
0in,keepaspectratio=false,width=2.2in,angle=-0,clip=false]{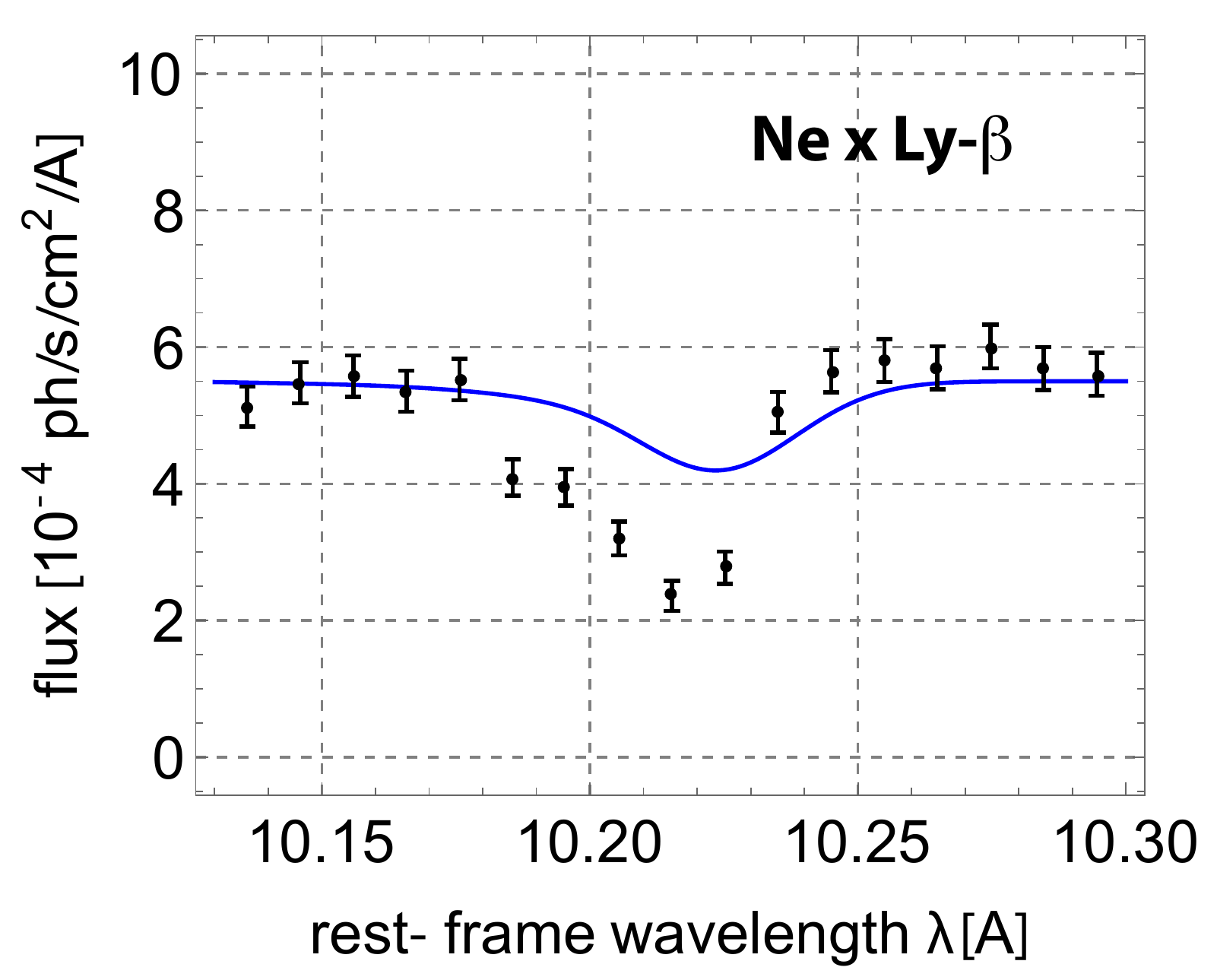}
\includegraphics[trim=0in 0in 0in
0in,keepaspectratio=false,width=2.2in,angle=-0,clip=false]{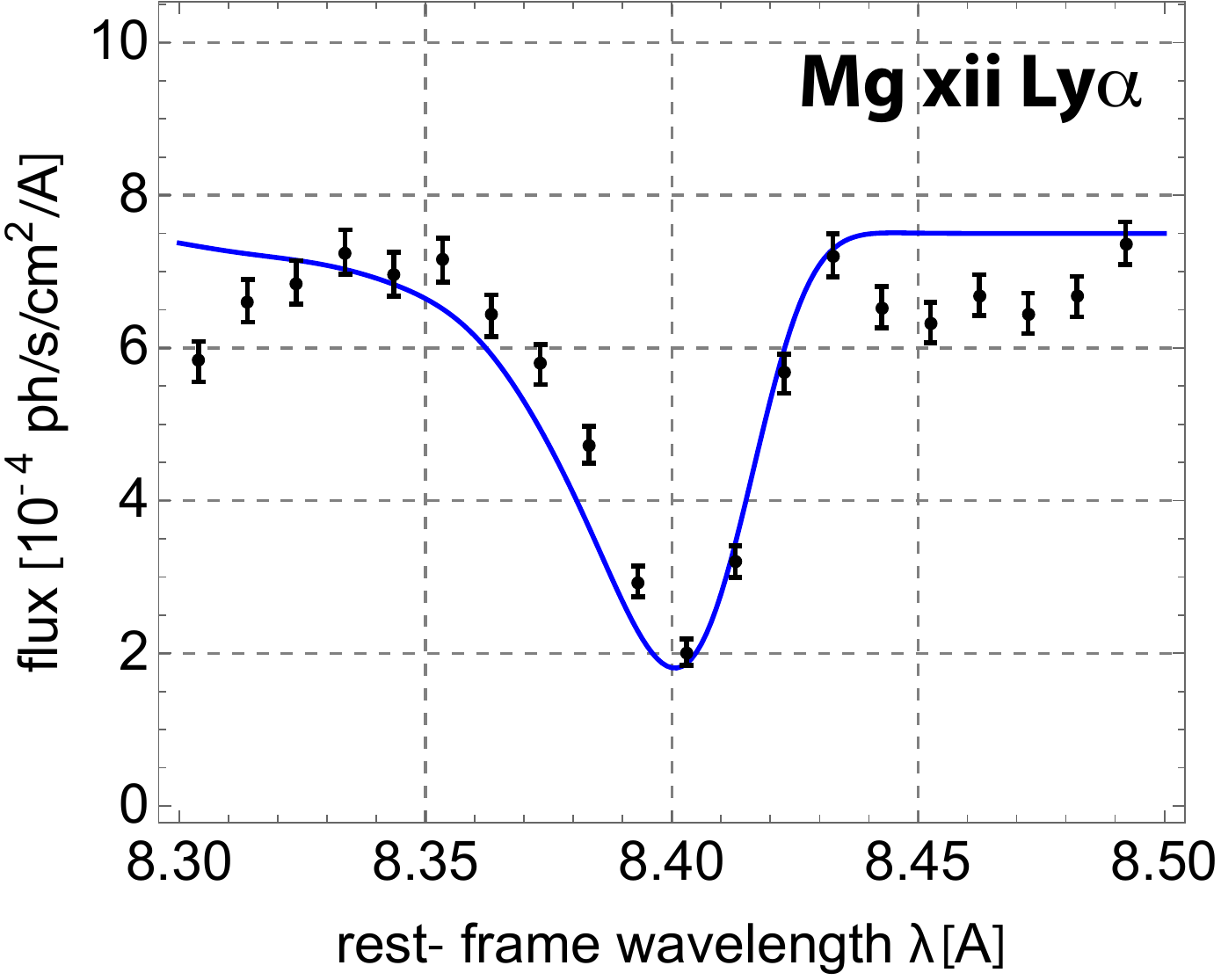}
\includegraphics[trim=0in 0in 0in
0in,keepaspectratio=false,width=2.2in,angle=-0,clip=false]{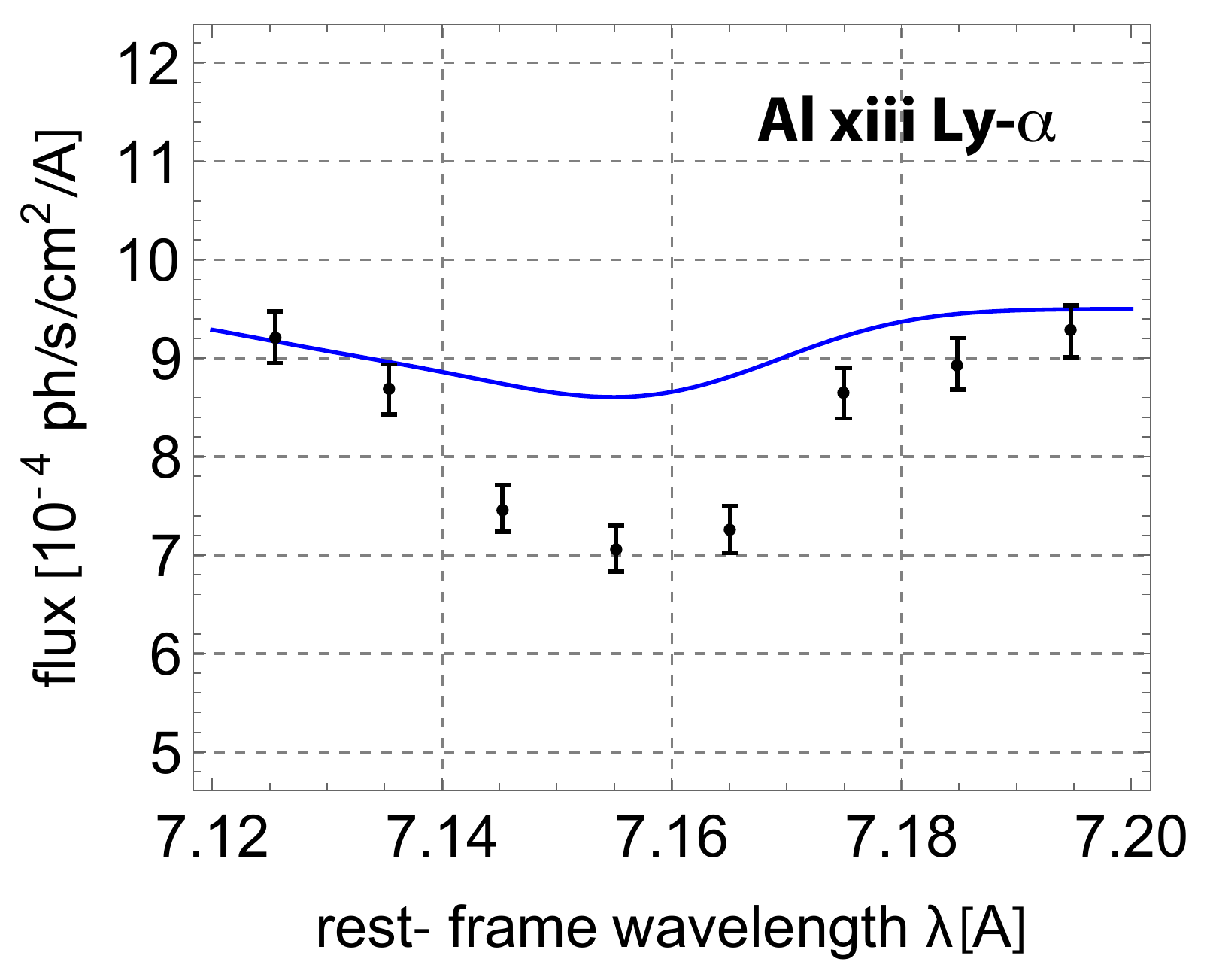}
\\
\includegraphics[trim=0in 0in 0in
0in,keepaspectratio=false,width=2.2in,angle=-0,clip=false]{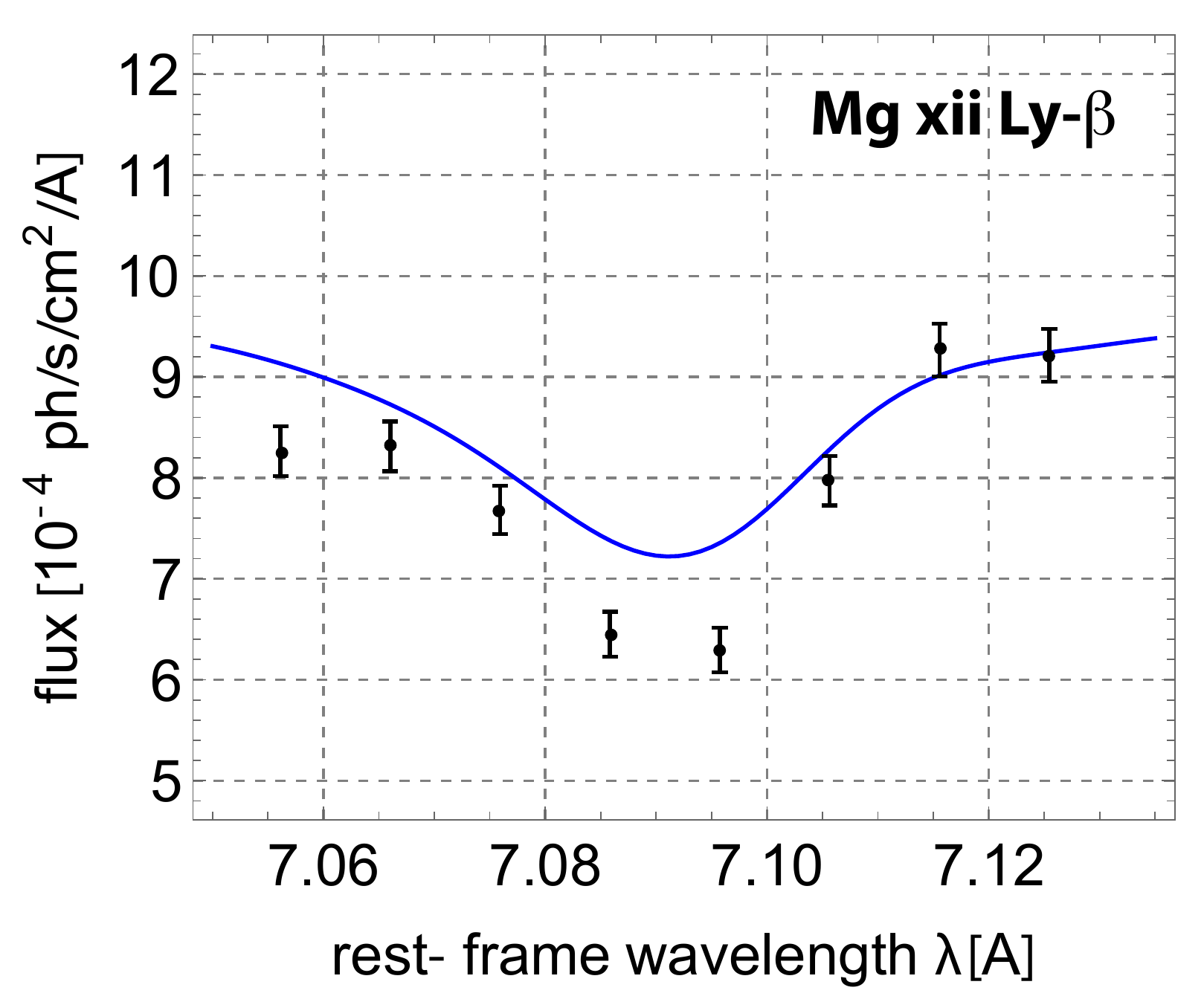}
\includegraphics[trim=0in 0in 0in
0in,keepaspectratio=false,width=2.2in,angle=-0,clip=false]{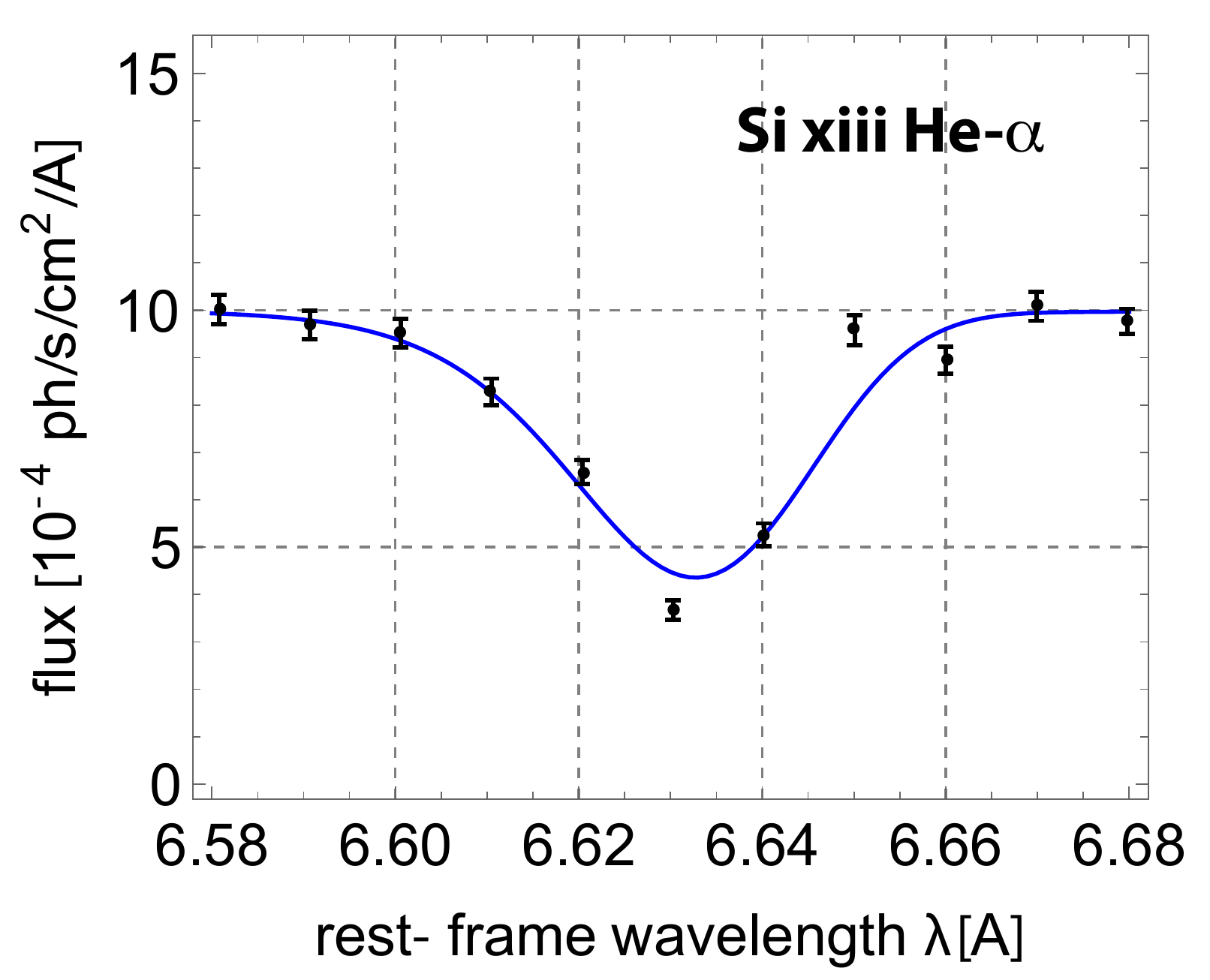}
\includegraphics[trim=0in 0in 0in
0in,keepaspectratio=false,width=2.2in,angle=-0,clip=false]{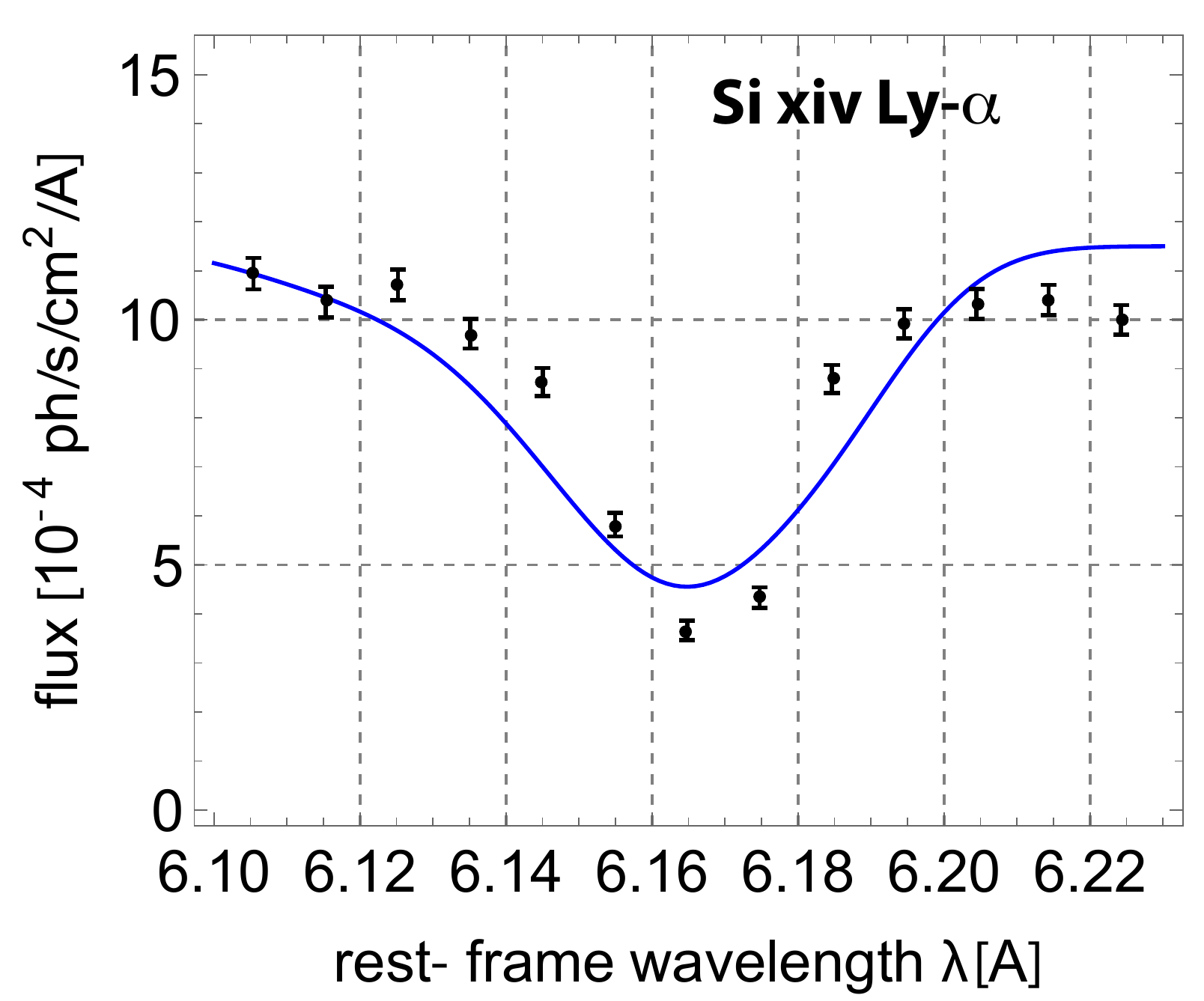}
\end{array}$
\end{center}
\caption{Sample of the best-fit spectra  calculated from the modeled MHD-wind (in blue lines)
with $p=1.15$, $\theta_{\rm obs}=44\deg$ and $n_{11} = 6.9$. See {\bf Tables~\ref{tab:tab2}
and \ref{tab:tab3}} for more details. }
\label{fig:spec1}
\end{figure}

Our investigation begins with calculating the structure of various magnetized wind
solutions of different global density profiles  that essentially determine  the
absorber's AMD. Specifically, we consider $p=1.29, 1.15, 1, 0.9, 0.8$ and $0.7$ consistent with the relevant range obtained in B09 for \obj. For photoionization
calculations with {\tt xstar}, we adopt the same ionizing SED of \obj\ as determined in
K01 (see their Tab.~4) by the LFZ continuum fitting, i.e. a composite spectrum of
multiple power-law components of different slopes between $0.2$ eV and $30$ keV, namely
$\Gamma=2.0$ (0.2 eV - 2 eV), $1.5$ (2 eV - 40 eV), 5.77 (40 eV - 0.1 keV) and 1.77
(0.1 keV - 30 keV).  We adopt the ionizing luminosity, $L_X = 3 \times 10^{43}$ erg~s$^{-1}$, in $1-1000$ Rydberg (see, e.g., K01 and K02) for photoionization calculations.
AGN SED is generally broad-band, covering wavelengths much broader than the X-rays. However, UV photons of energy below 1 Rydberg (13.6 eV, the threshold energy for ionizing hydrogen) do not contribute to photoionization of the wind, while harder X-ray photons of energy higher than 1000 Rydberg (13.6 keV) typically have much lower flux (compared to that of soft X-rays) and do not make a significant impact on the overall photoionization process. The broader disk SED photons emitted near the black hole ($r \gsim r_o$) contribute to the cooling of the ionized gas and those have been included in our calculations.

This SED is injected at the coordinate origin as an irradiating
source, mimicking the compact point source of an X-ray corona. As mentioned, a given
SED will uniquely determine the ionization structure of the wind that ``breaks" the
otherwise mass-invariant spatial nature of the MHD-driven X-ray absorbers.
To show a predictive power of our models, {\bf Figure~\ref{fig:amd}} shows, as one of many (typically 12-15) template calculations, a synthetic AMD for a number of ions computed with $p=1.15$ wind for $\theta_{\rm obs}=40\deg$ and $n_{11} = 13.6$ where $n_o \equiv n_{11} 10^{11}$ cm$^{-3}$. As seen, the peak column of a given ion increases with ionization parameter $\xi$ (also with decreasing distance $r$) and the slope of this AMD ($=0.176$ from eqn.~(\ref{eq:amd2}); dashed line), consistent with that derived in B09, is set by the wind density slope of $p=1.15$ assuming the solar abundances. The LoS wind velocity scales as Keplerian, $r^{-1/2}$, in this model.

\begin{figure}[ht]% ------------------------------------- Figure~3 (2 continued)
\begin{center}$
\begin{array}{ccc}
%\includegraphics[trim=0in 0in 0in
%0in,keepaspectratio=false,width=4.0in,angle=-0,clip=false]{bestfit_r115_el.pdf}
\includegraphics[trim=0in 0in 0in
0in,keepaspectratio=false,width=2.2in,angle=-0,clip=false]{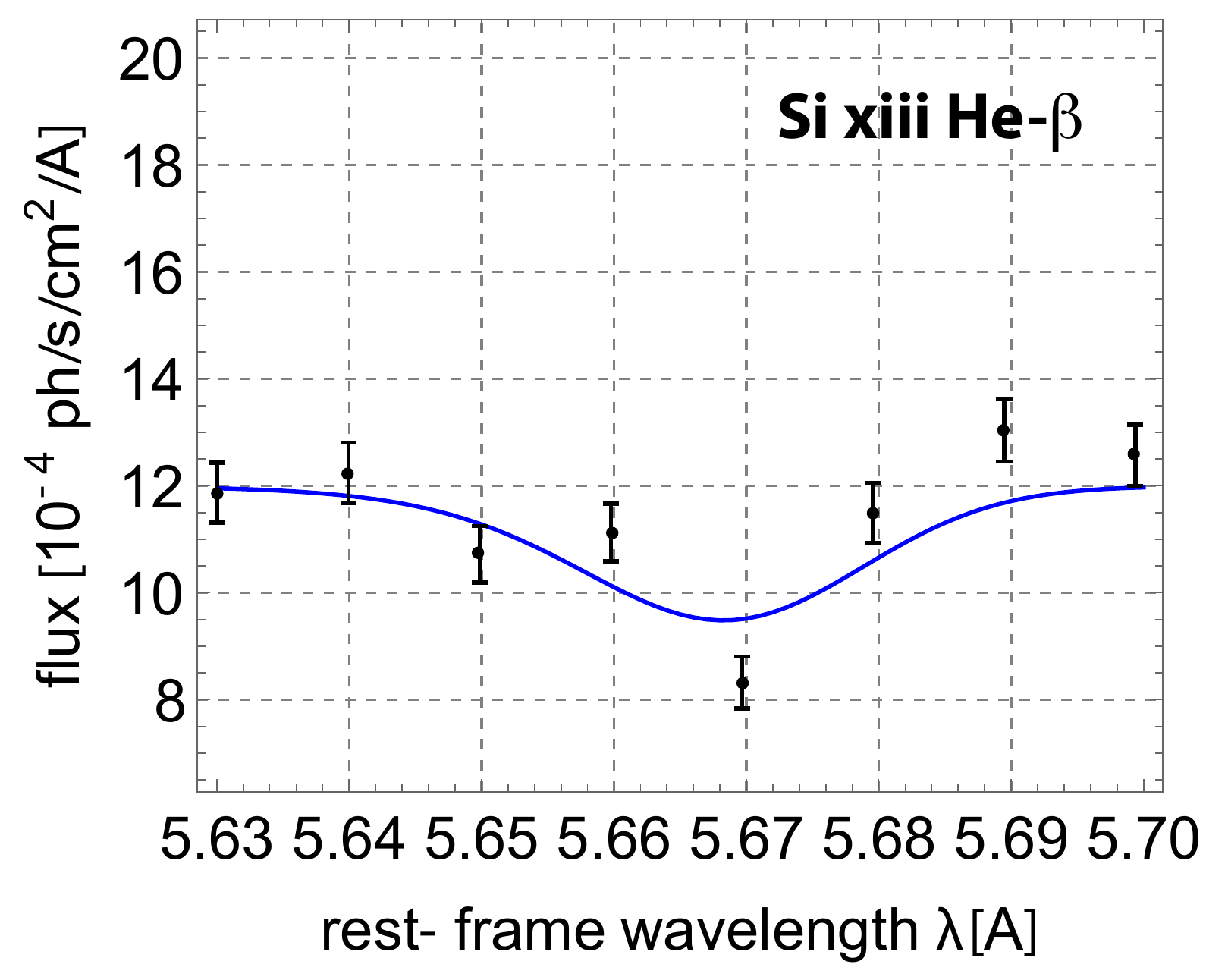}
\includegraphics[trim=0in 0in 0in
0in,keepaspectratio=false,width=2.2in,angle=-0,clip=false]{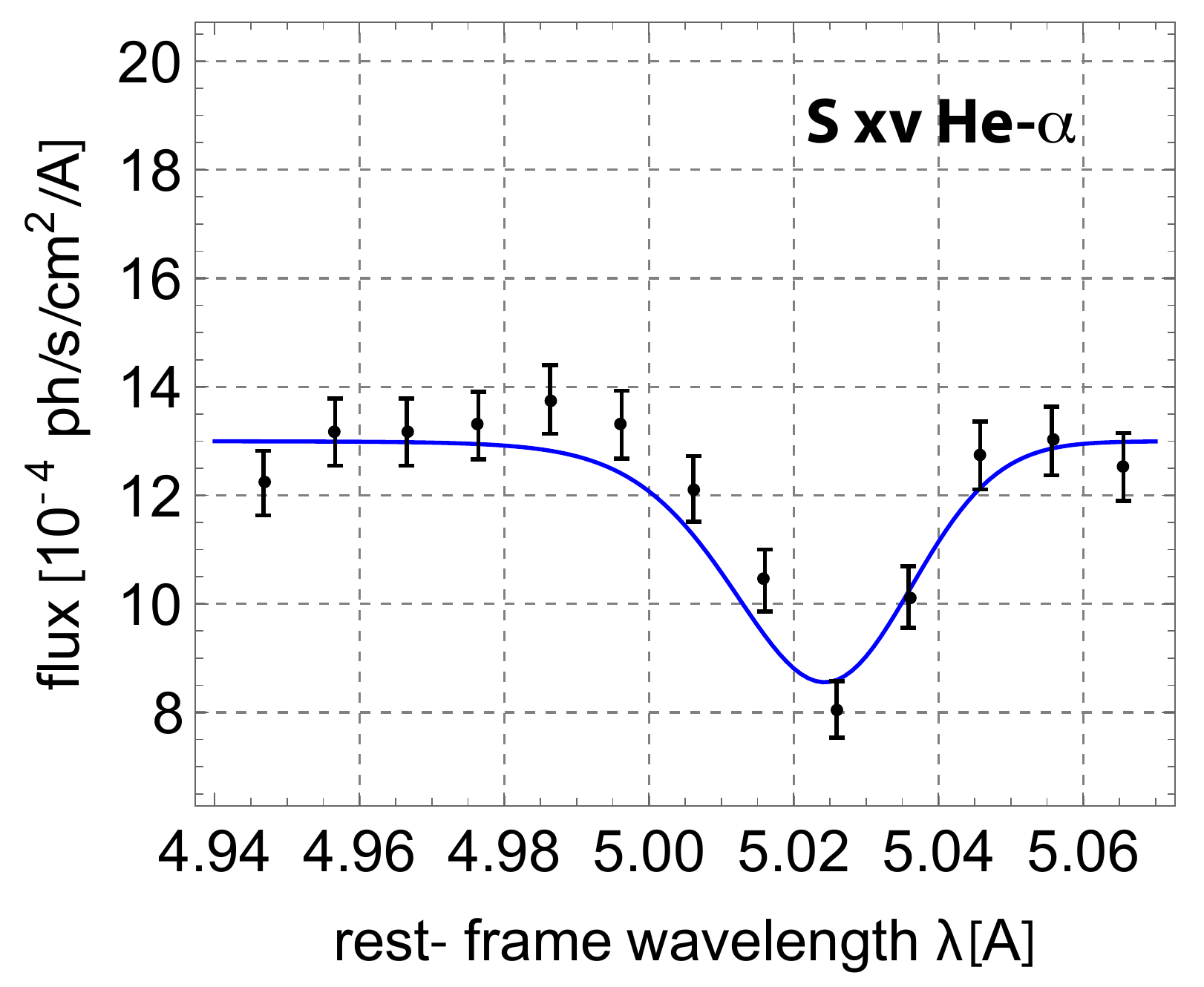}
\includegraphics[trim=0in 0in 0in
0in,keepaspectratio=false,width=2.2in,angle=-0,clip=false]{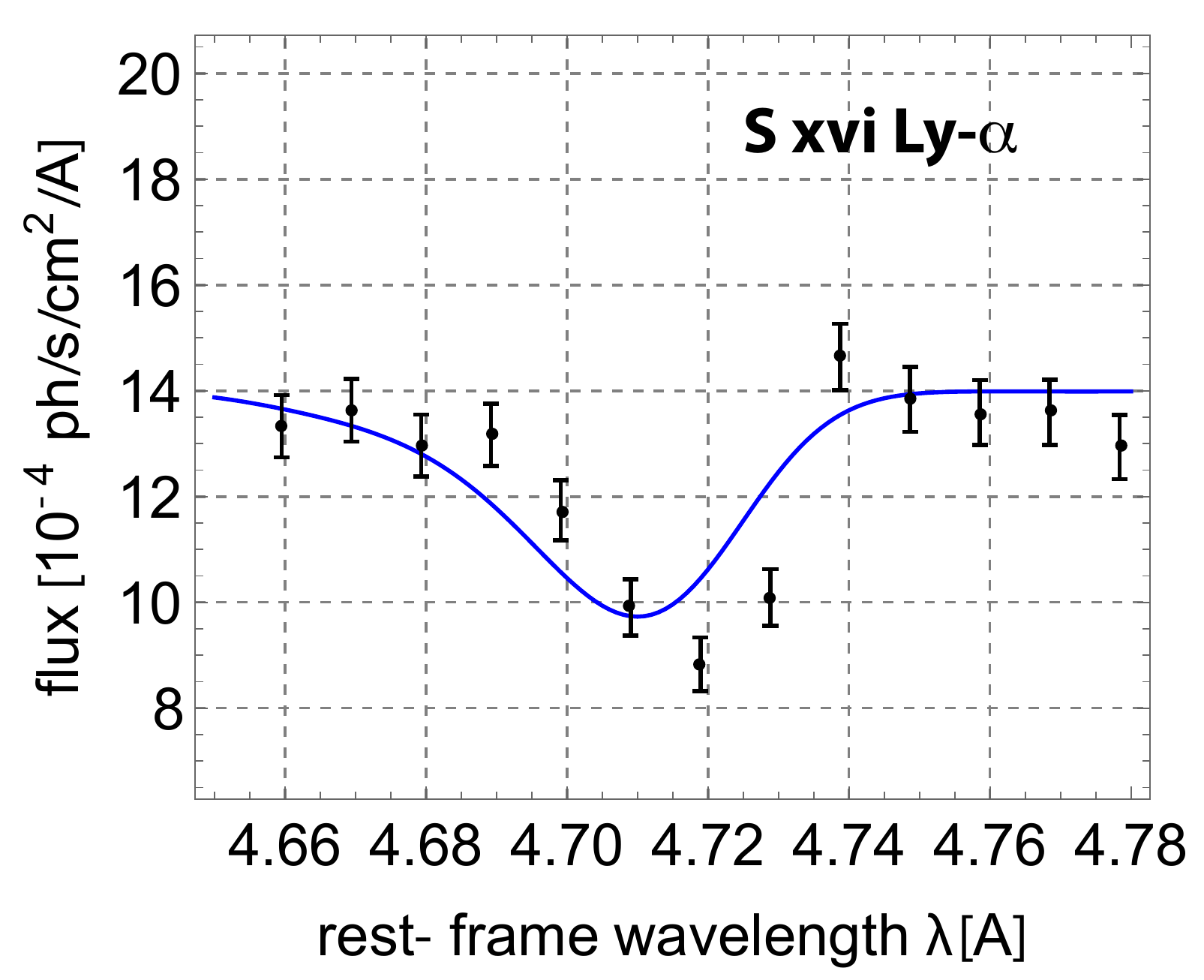}
\\
\includegraphics[trim=0in 0in 0in
0in,keepaspectratio=false,width=2.2in,angle=-0,clip=false]{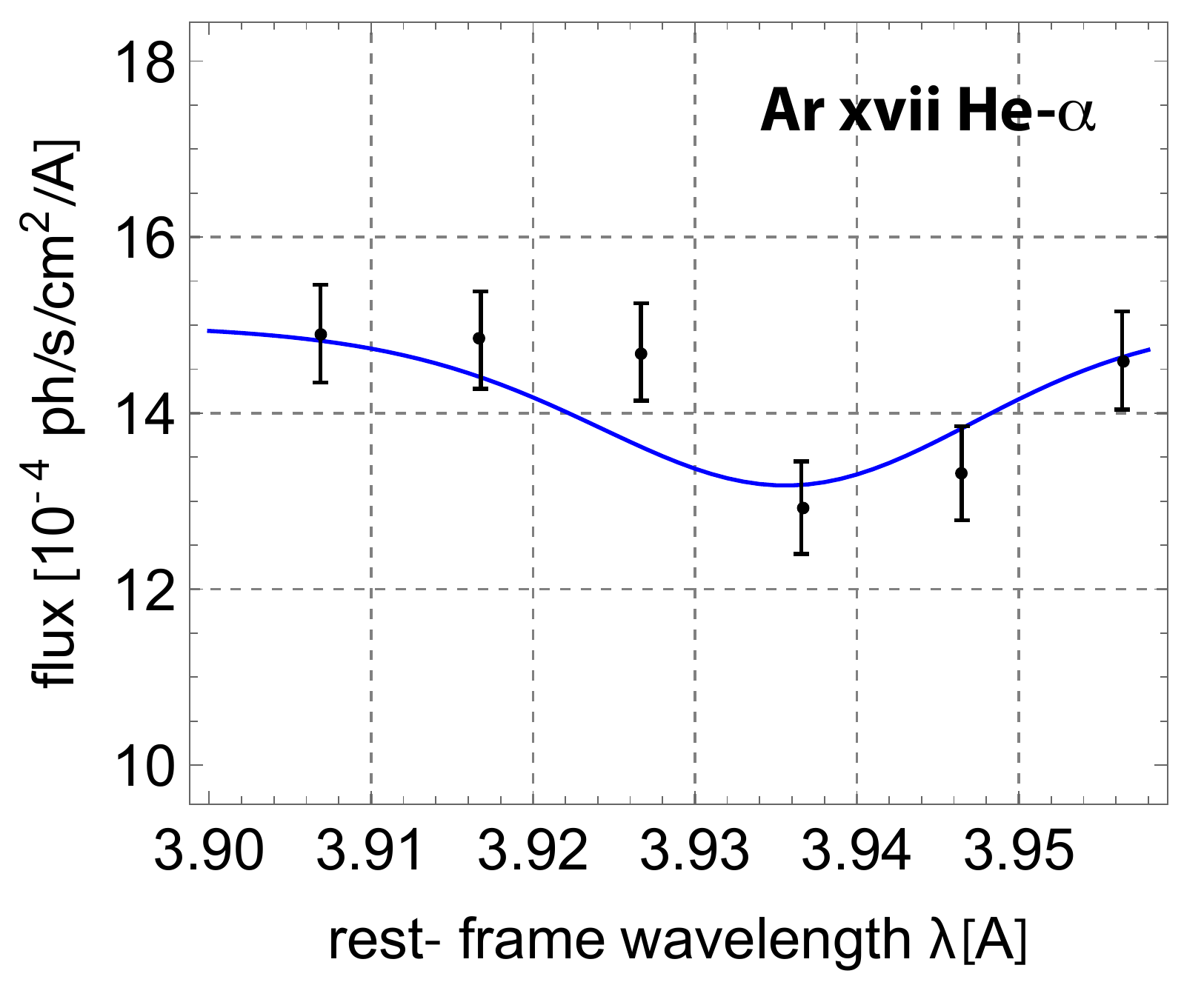}
\includegraphics[trim=0in 0in 0in
0in,keepaspectratio=false,width=2.2in,angle=-0,clip=false]{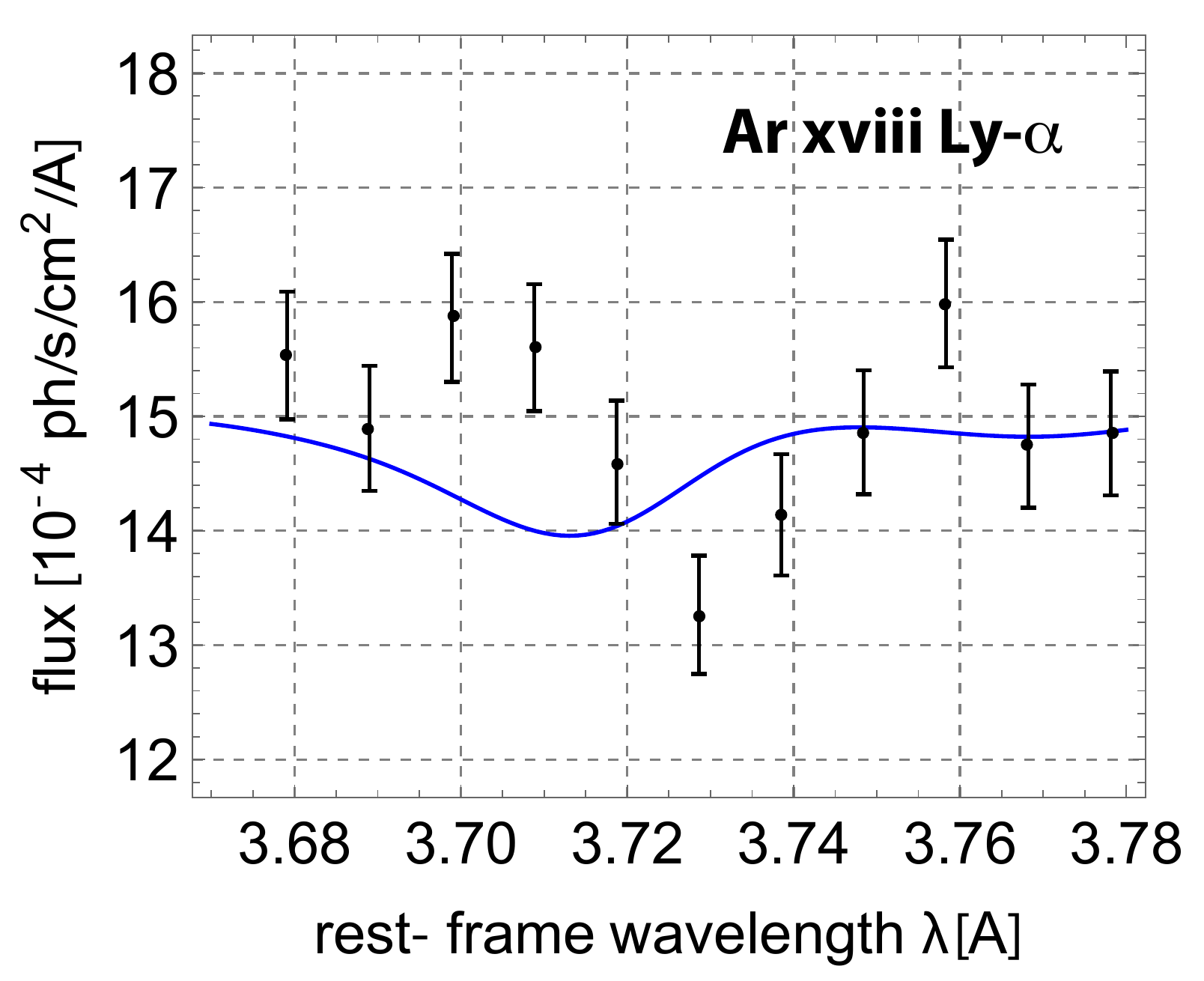}
\includegraphics[trim=0in 0in 0in
0in,keepaspectratio=false,width=2.2in,angle=-0,clip=false]{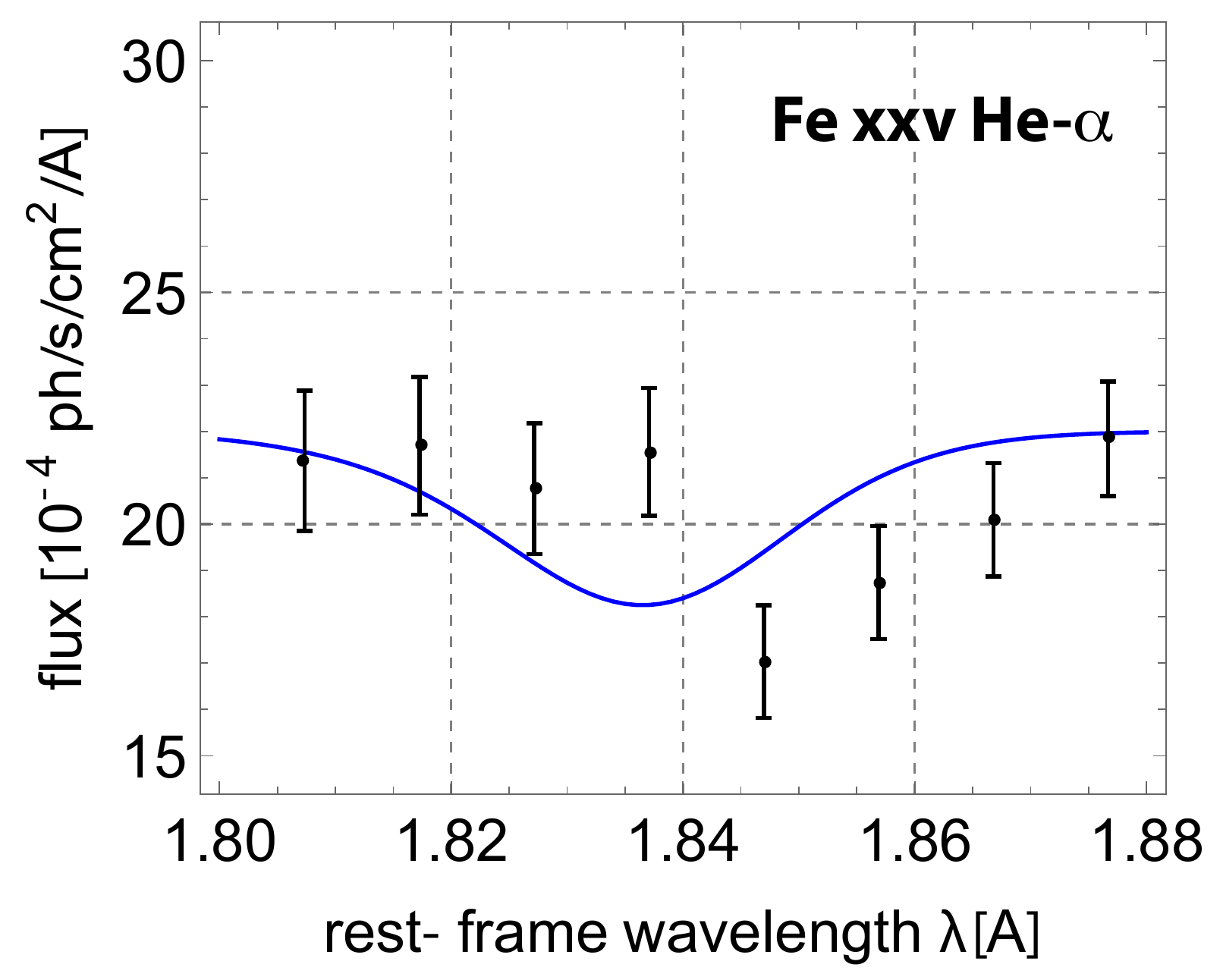}
\end{array}$
\end{center}
\caption{Sample of the best-fit spectra calculated from the modeled MHD-wind (in blue lines)
with $p=1.15$, $\theta_{\rm obs}=44\deg$ and $n_{11} = 6.9$. See {\bf Tables~\ref{tab:tab2}
and \ref{tab:tab3}} for more details. }
\label{fig:spec2}
\end{figure}

In our investigation, 18 absorption lines (see {\bf Tables~\ref{tab:tab2}-\ref{tab:tab3} and Figs.~\ref{fig:spec1}-\ref{fig:spec2}} which will be discussed later more in \S 4) are modeled systematically and globally, in the sense that
they are all computed within the confines of a single, continuous disk wind;
this can pose tight restrictions on the model, since, in such a global model, we cannot
afford to adjust individual absorber parameters independently to obtain good fits for
specific features. As a result, for a given density distribution $p$ (obtained by
spectral fitting and AMD), our model has only two free parameters: (1) The wind density
normalization $n_{o}$ at the innermost launching radius at $r \gsim r_S$ on the disk surface and (2) The
disk inclination angle $\theta_{\rm obs}$. Elemental abundances could be adjusted
individually, but solar abundances are assumed here. Finally,
considering that the observed  spectrum also exhibits a number of prominent emission
lines, especially towards longer wavelengths, we have added, as needed, emission
features to our model continuum spectrum of the equivalent width (EW) derived in K02.
%that makes it very challenging to even accurately assess the local
%continuum level, our analysis result could be influenced. Finally, since our present
%investigation is not motivated by obtaining the best $\chi^2$-statistical results but
%rather driven by a hypothesis that the observed WAs can be originating from a
%stratified magnetic wind, we feel that the absolute values of $\chi^2$ from our
%analysis below do not reflect much of our physical context, although we will discuss
%and statistically address various models for their comparative relevance.

%
%

%%------------------------------- Table~1
%\begin{deluxetable}{l c ccc}
%\tablecaption{Characteristics of Fiducial GRMHD Plasma Accretion \label{tab:tab1}}
%\tabletypesize{\small}
%%\rotate
%% ^ this control sequence will rotate the table, which can be useful when your table gets rather wide.
%\tablecolumns{10}
%\tablewidth{0pt}
%\tablehead{
%	\colhead{Model Parameter} & \colhead{Description} &
%	\multicolumn{1}{c}{ $a/M$} \\
%	\cline{3-5} \\
%	\colhead{} & \colhead{} &
%	\colhead{-0.5} & \colhead{0} & \colhead{0.5}
%}
%\startdata
%$E$ & Energy & 6.1 & 6.1 & 6.1  \\
%$L/E$ & Specific angular momentum & 2.1 & 2.3 & 3.96  \\
%$\Omega_F$ & Angular velocity of field line & 0.02725 & 0.08334 & 0.2333  \\
%$4\pi \eta$ & Scaled accretion energy  & 0.0082 & 0.006 & 0.005 \\  \hline
%%$r_{\rm inj}$ & Injection point & \\
%$r^{\rm out}_A/r_g$ & Outer \Alfven radius & 7.01 & 3.28 & 2.82 \\
%$r^{\rm out}_F/r_g$ & Outer fast radius & 3.52 & 3.01 & 2.49 \\
%$r_{\rm sh}/r_g$ & Shock location & 2.23 & 2.70 & 1.99 \\
%$r^{\rm in}_F/r_g$ & Inner fast radius & 2.09 & 2.41 & 1.90 \\
%$r_H/r_g$ & Event horizon & 1.86  & 2.0 & 1.86 \\
%$\Theta_e(r=r_{\rm sh})$ & Electron energy  & 0.358 & 0.199 & 0.285 \\
%\enddata \\
%%\tablenotetext{a}{Table notes can be useful for giving further explanation to quantities and adding notes about the table contents.}
%\vspace{0.1cm}
%Note: Superscripts ``in" and ``out" respectively denote those radii for the ``downstream" and ``upstream" plasma.
%\vspace*{0.2cm}
%\end{deluxetable}

\clearpage

%\section{Results}

%------------------------------- Table~1
\begin{deluxetable}{l|c}
\tabletypesize{\small} \tablecaption{Primary Grid of MHD-Wind Model Parameters } \tablewidth{0pt}
\tablehead{Primary Parameter & Value}
\startdata
Wind density slope $p$ & $0.7, 0.8, 0.9, 1.0, 1.15, 1.29$ \\
Inclination angle $\theta$ [degrees] & $30\deg, 40\deg, 50\deg$ \\
Wind density normalization $n_{11}$$^{\dagger}$  & 0.1 - 100 \\
%Downstream Elecetron Energy $kT_e$ & See \S 3.3 \\
%Thickness  $H \equiv h/r$  & $0.1, 0.5, 1$   \\
%Mass-Accretion Rate $\dot{m} \equiv \dot{M}/\dot{M}_{E}$ & $0.1, 0.5, 1$  \\
%Accreting Plasma  $r_{\rm sh}/r_g$  & $2?, 3?$   \\
\enddata
\vspace{0.05in}
%\begin{flushleft}
We assume $M = 3 \times 10^7 \Msun$ \citep{Peterson04} and $L_X = 3 \times 10^{43}$ erg~s$^{-1}$ (K02).
\\
$^\dagger$ Wind density normalization in units of $10^{11}$ cm$^{-3}$.
%\end{flushleft}
\label{tab:tab1}
\end{deluxetable}

\clearpage

First, we try to narrow down an optimal  range of the model parameters by crudely
exploring the parameter space spanned by $(n_{11}, \theta_{\rm obs})$ for a series of
wind density structure given by $p$ where $n_{11}$ is defined as $n_o \equiv n_{11} 10^{11}$ cm$^{-3}$.
The likely value of $p$ is successfully constrained
within the error set by B09 and by \citet{Laha14}. Following this preliminary
investigation, a primary grid of models is determined as shown in {\bf
Table~\ref{tab:tab1}}. A more thorough analysis is then conducted for these template
models to compute in addition the physical parameters of individual ions (e.g. distance
$r$, column $N_H$, number density $n$, velocity $v_{\rm out}$, ionization parameter
$\xi$, plasma temperature $T$); this is done for the major spectral signatures, i.e.
for a total of 18 transition lines from major H/He-like ions; i.e. Fe, Ar, S, Si, Mg, Al, Ne, O and N
of relatively large EW found in K02 over the $\sim 1\aa - 20\aa$ band, as
they present the most significant indicators of the bestfit global wind solution.
%Fitting results are then examined with
%the standard $\chi^2$ per degree of freedom (dof) statistic and comparison of the
%model/observed EW as discussed below. This analysis method is similar to that in F17.

\section{Results}

\begin{figure}[ht]% ------------------------------------- Figure~4
\begin{center}$ \begin{array}{ccc}
\includegraphics[trim=0in 0in 0in 0in,keepaspectratio=false,width=2.1in,angle=-0,clip=false]{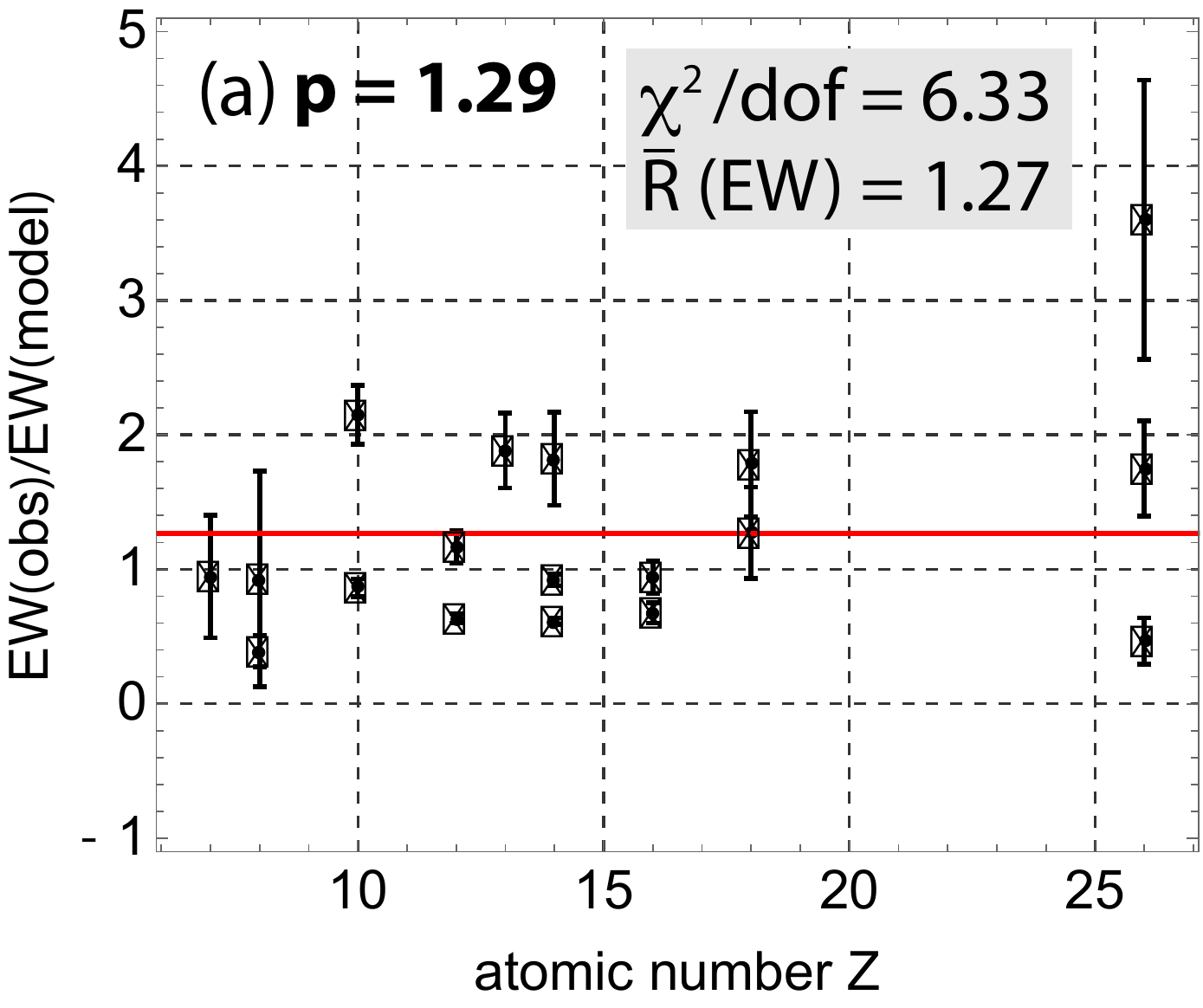}
\includegraphics[trim=0in 0in 0in 0in,keepaspectratio=false,width=2.1in,angle=-0,clip=false]{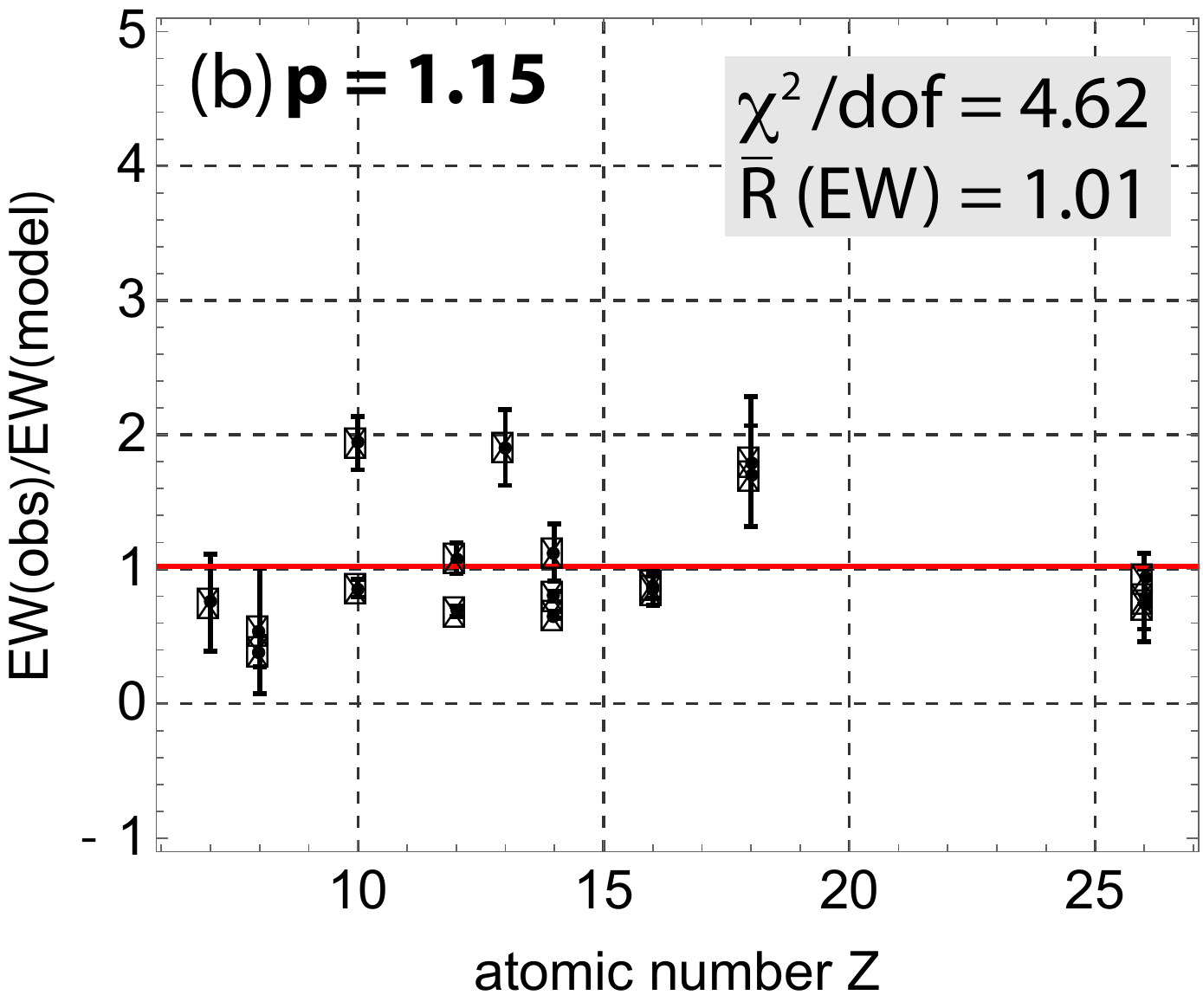}
\includegraphics[trim=0in 0in 0in 0in,keepaspectratio=false,width=2.1in,angle=-0,clip=false]{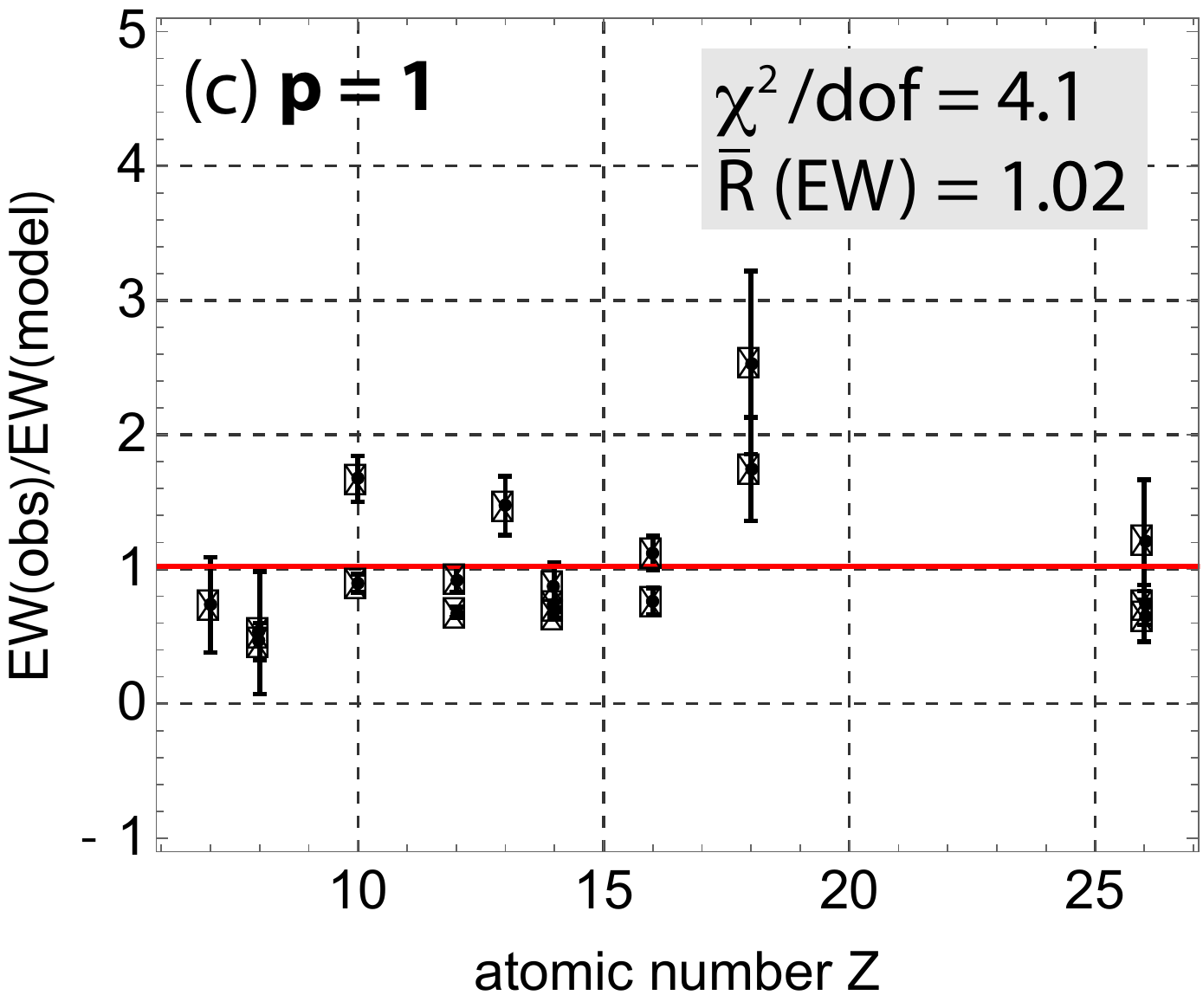} \\
\includegraphics[trim=0in 0in 0in 0in,keepaspectratio=false,width=2.1in,angle=-0,clip=false]{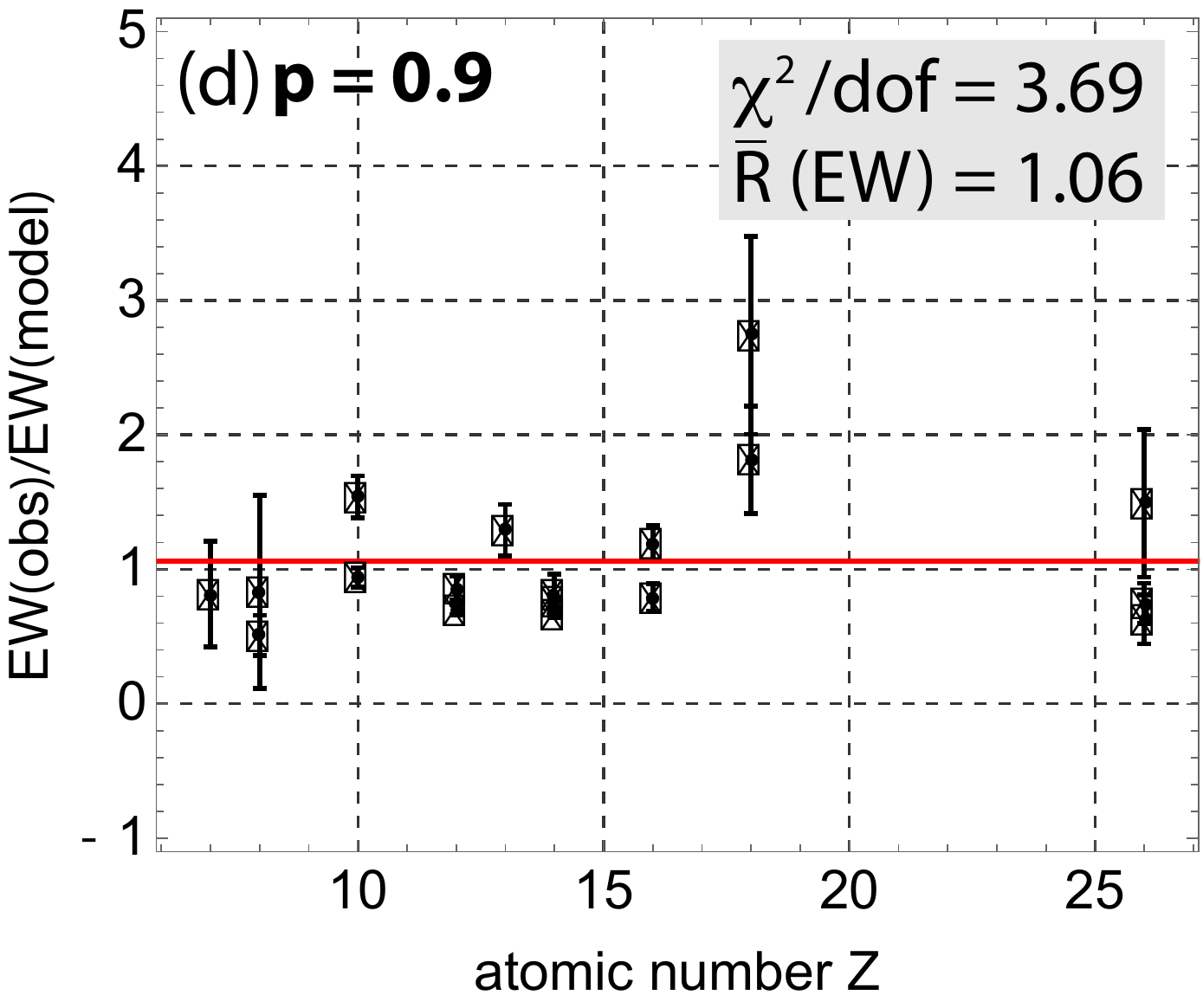}
\includegraphics[trim=0in 0in 0in 0in,keepaspectratio=false,width=2.1in,angle=-0,clip=false]{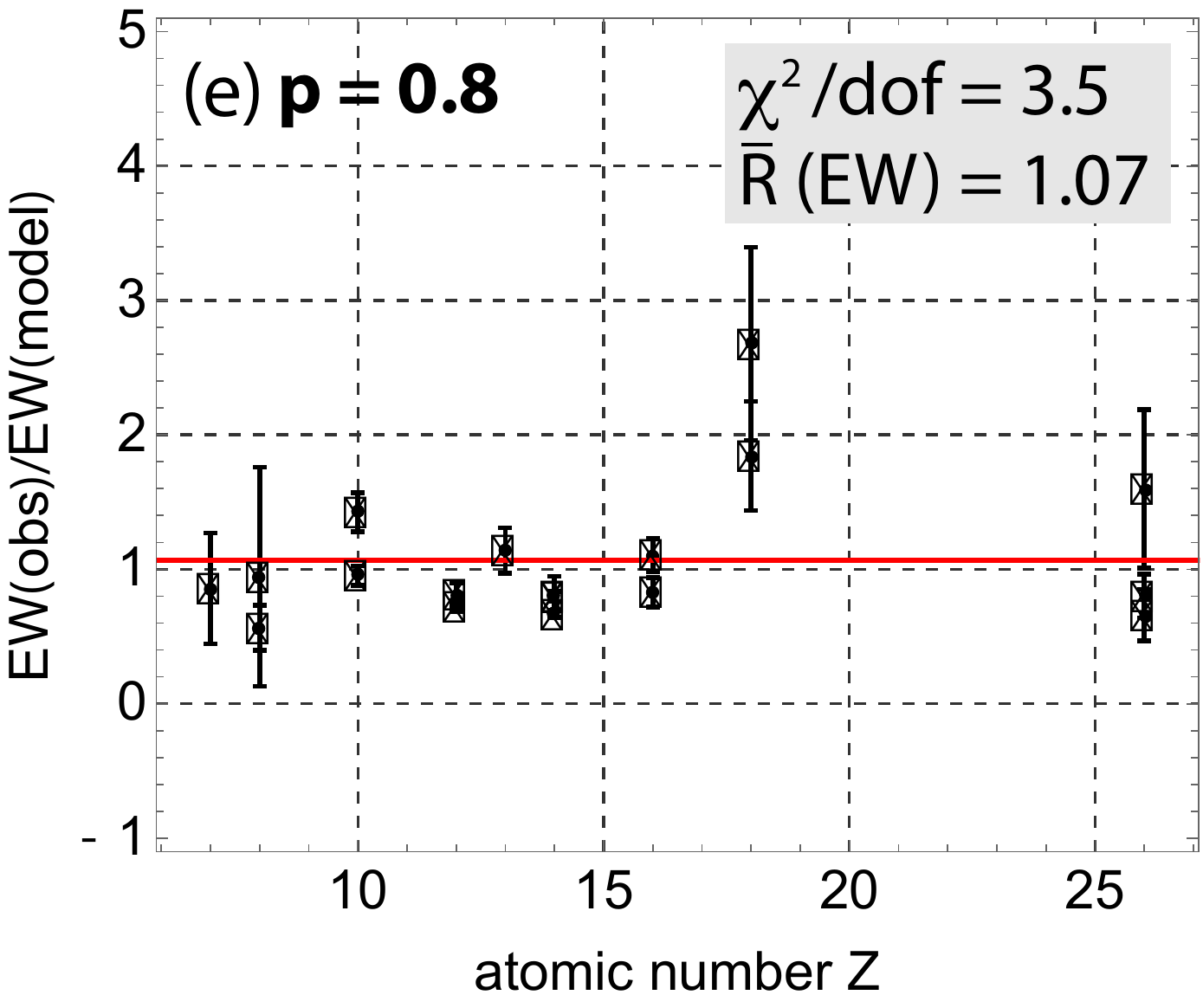}
\includegraphics[trim=0in 0in 0in 0in,keepaspectratio=false,width=2.1in,angle=-0,clip=false]{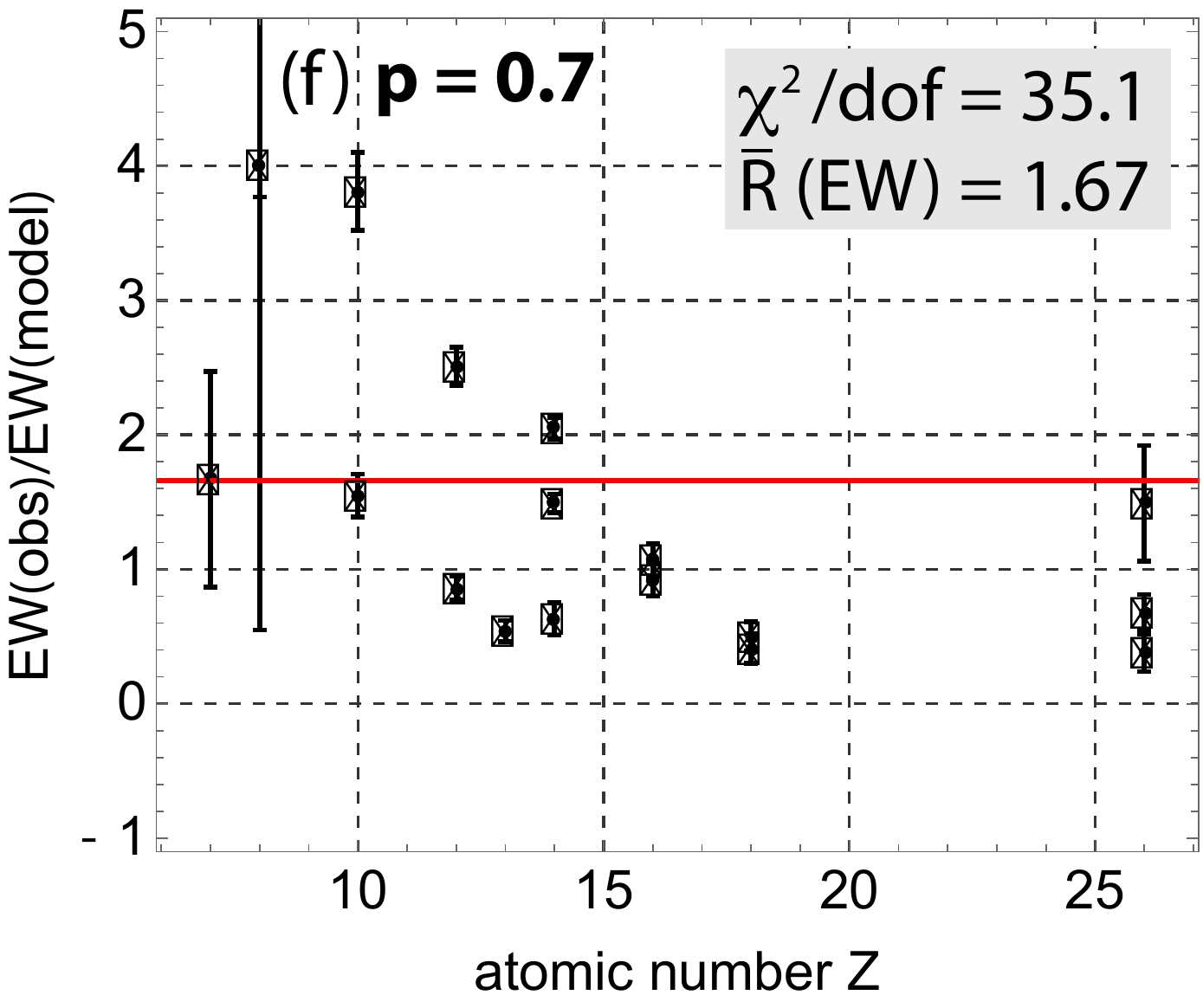}
\end{array}$
\end{center}
\caption{EW ratios, $R_{\rm EW} \equiv$ EW(obs)/EW(model), between the model and observation for the same 18
ions shown in {\bf Table~\ref{tab:tab2}} derived from the bestfit models for various wind
density profiles; (a) $p=1.29$, (b) 1.15, (c) 1.0, (d) 0.9, (e) 0.8 and (f) 0.7 along
with its {\it mean} value, $\bar{R}_{\rm EW}$, (shown in red lines). Also shown are the
derived $\chi^2$/dof values for each bestfit model from 18 ions in {\bf Table~\ref{tab:tab2}}.  As seen,
the $p=1.0 - 1.15$ wind is physically more favored at a statistically significant level. }
\label{fig:EW}
\end{figure}

To assess  quantitatively the  model's statistical significance,
we create a large  number (typically 12-15) of template wind solutions with different sets of $(n_{11}, \theta_{\rm obs})$ where $0.1 \le n_{11} \le 100$ and $30\deg \le \theta_{\rm obs} \le 60\deg$ for a given $p$ within the relevant range inferred from B09, also as shown in {\bf Table~1}.
For each such pair of $(n_{11},\theta_{\rm obs})$, we compute the predominantly strong absorption line profiles and calculate the $\chi^2$/dof from the ensemble of the 18
absorption lines. Minimization of $\chi^2$ for each density profile yields the bestfit spectrum solution and we calculate the EW for each transition line both from our wind model and data. Besides the value of $\chi^2$, we also calculate a ratio, as another useful proxy for goodness-of-fit, $R_{\rm EW}  \equiv {\rm EW(obs)/EW(model)}$, of the observed to the model EW for each ion. In {\bf Figure~\ref{fig:EW}} we plot the ratios $R_{\rm EW}$ as a function of atomic number $Z$ along with their mean value $\bar R_{\rm EW}$ (shown by red lines). The inset of each panel denotes the resulting value of $\bar R_{\rm EW}$ along with the corresponding $\chi^2$/dof values for the specific value of $p$.
Assuming that the wind characteristics vary monotonically with the wind density parameter $p$, it is implied here that the {\it global} bestfit solution spanned by the $(n_{11}, \theta_{\rm obs}; p)$-space is found somewhere around $p=1.15$, which is in fact consistent with the result from B09.

\begin{figure}[h]% ------------------------------------- Figure~5
\begin{center}$ \begin{array}{cc}
\includegraphics[trim=0in 0in 0in 0in,keepaspectratio=false,width=3.05in,angle=-0,clip=false]{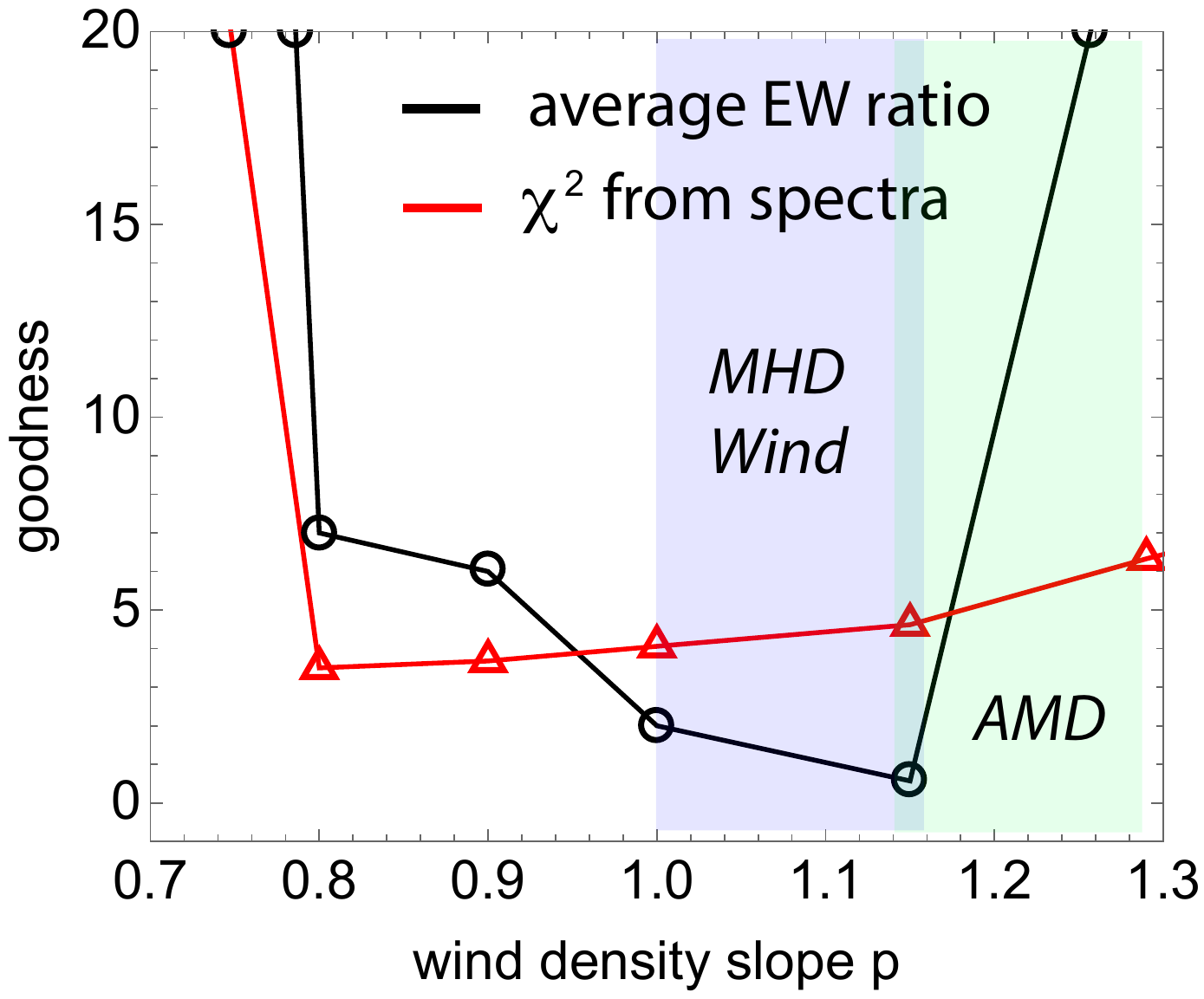}
\includegraphics[trim=0in 0in 0in
0in,keepaspectratio=false,width=3.4in,angle=-0,clip=false]{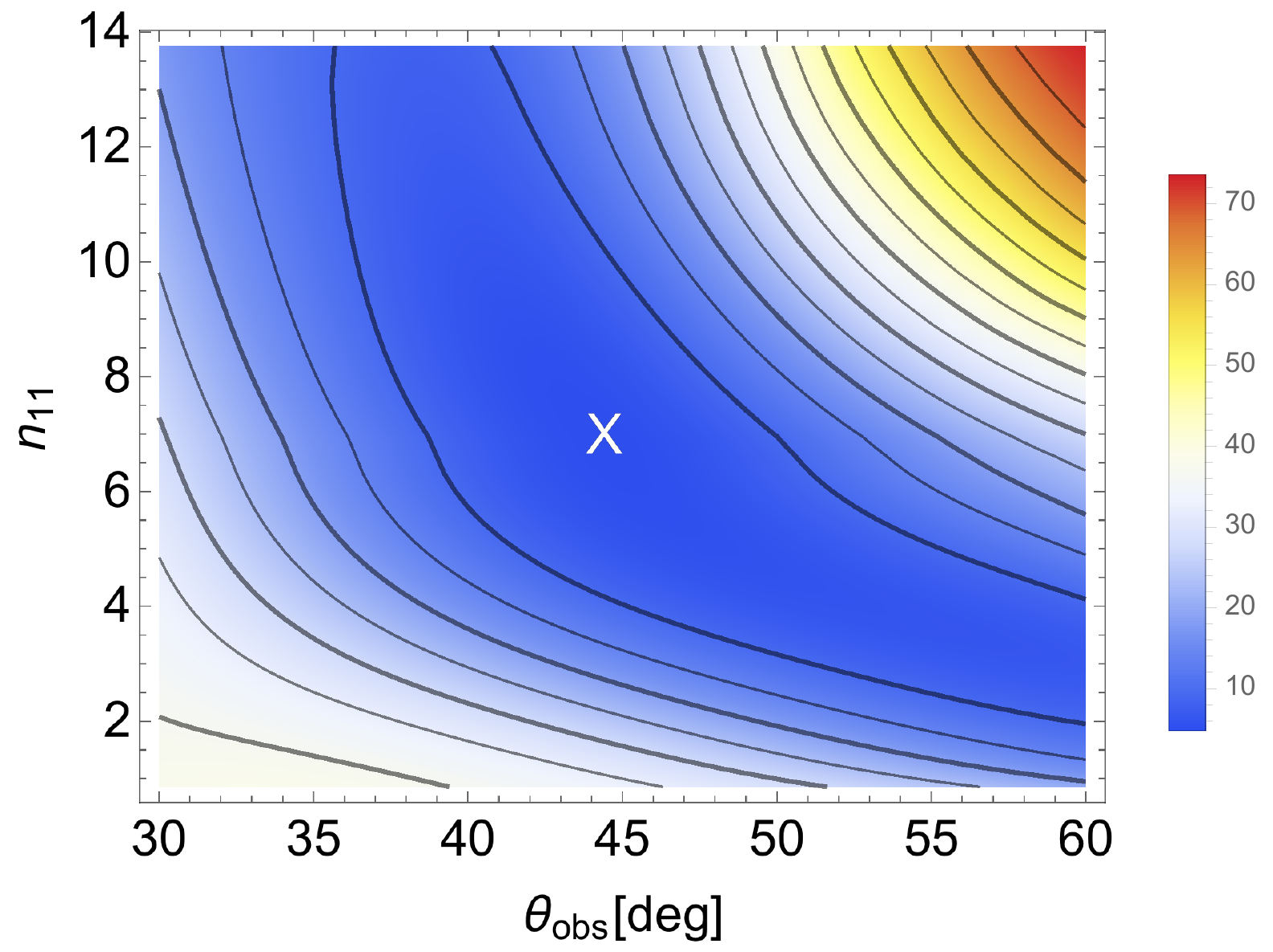}
\end{array}$
\end{center}
\caption{(a) Goodness-of-fit assessed by $\chi^2$-statistics (red triangles) and the {\it mean}
EW ratios (dark circles), $\bar{R}_{\rm EW} \equiv 100(\bar{R}-1)$, i.e. its \% deviation from a perfect match,
corresponding to {\bf Figure~\ref{fig:EW}} for different density slope $p$ considered in this work.
The light blue band indicates the most likely values based on the combination of the two
constraints. Note that the points for $p=0.7$ are off the range as similarly shown in {\bf Figure~\ref{fig:EW}}. The most likely range from our bestfit (labeled as {\it MHD Wind}) and the AMD in B09 (labeled as {\it AMD}) are respectively denoted by shaded regions.
(b) A contour plot of $\chi^2$/dof values (color-coded) corresponding to the model parameters $(n_{\rm 11}, \theta_{\rm obs})$ from the
$p=1.15$ wind with its bestfit solution $\theta=44\deg$ and $n_{11}= 6.9$ (white X).
%For the latter measure, we normalized $\bar{R}$(EW) to its deviation from the perfect
%match in \% here. Also shown are likely ranges of the slope  inferred from the bestfit
%MHD-wind model (labeled as {\it MHD Wind} in blue region) and that derived from AMD
%analysis in \citet{B09} (labeled as {\it AMD} in green region), respectively. .
}
\label{fig:goodness}
\end{figure}

To better illustrate the goodness-of-fit in our spectral analysis, we plot in
{\bf Figure~\ref{fig:goodness}a } the values of $100(\bar{R}_{\rm EW} -1)$ (open circles with line) and the corresponding $\chi^2$/dof values (red triangles with line) as a function of the value $p$.
We do see that the $\chi^2$ statistics is rather insensitive to the value of $0.8 \lesssim p \lesssim 1.15$, but increases steeply outside this range. It should be noted here that $\chi^2$ statistics in this dataset is predominantly controlled by a small number of stronger  lines with very small error bars produced around $\xi \sim 2.5$. As long as the choice of a pair of $(n_{11},\theta_{\rm obs})$ produces the correct value of column density at the given value of $\xi$, the model gives a similarly small $\chi^2$/dof value. However, it fails here at higher values of $\xi$ because not many high-Z elements are detected with large significance in this data. On the other hand,
the EW ratios $\bar{R}_{\rm EW}$ fare better in that respect, as they provide an alternative to the
AMD covering a larger $\xi$-range.
From the combination of the two proxies here, we see again
that $p$ in the range $\sim 1 - 1.15$ provides the most satisfactory fit to the data.
The most likely range from our bestfit (labeled as {\it MHD Wind}) and the AMD in B09 (labeled as {\it AMD}) are respectively denoted by shaded regions.
In the
remainder of this paper, therefore, we choose the value $p=1.15$ as the bestfit wind solution from a global perspective. For $p=1.15$, we also present in {\bf Figure~\ref{fig:goodness}b} a
contour map of $\chi^2$/dof (color-coded as shown) in the $\theta_{\rm obs} - n_{11}$ plane to obtain the bestfit solution of $\theta \simeq 44 \deg$ and $n_{11} \simeq 6.9$ (denoted by cross).

Focused on the obtained global bestfit solution, we have presented earlier in {\bf Figures~\ref{fig:spec1}-\ref{fig:spec2}} the bestfit profiles of the 18 lines of
\textbf{Table~\ref{tab:tab2}} with $p=1.15$, $\theta_{\rm
obs}=44\deg$ and $n_{11}=6.9$ (i.e. $n_o = 6.9 \times 10^{11}$ cm$^{-3}$) overlaid on
the data. Note that our spectral calculations employ the Voigt profile as described
above, with the wind shear as the only line broadening process, eschewing the use of an
artificial turbulent velocity, customarily employed in similar analyses.
This set of wind parameters is statistically favored among the models considered here
(as demonstrated in {\bf Fig.~\ref{fig:goodness}}).
% yielding $\chi^2$/dof=5.59.
As also seen in {\bf Figures~\ref{fig:spec1} and \ref{fig:spec2}}, the global bestfit model (i.e. $p=1.15$ wind) is generally
consistent with  most significant absorber transitions, both in line depth and shape
(the \nex \,Ly$\beta$ feature may involve line blending, not considered here).  As
discussed elsewhere in this paper, the \fexxv\ He$\alpha$ line, typically strong in the
AGN absorption spectra \citep{Tombesi13},  is rather weak in this object, in agreement
with our model, lending additional support to it.

\begin{deluxetable}{lc|ccccc}
\tabletypesize{\small}
\tablecaption{Line Measurements and Bestfit Model for 18 Absorbers } \tablewidth{0pt}
\tablehead{\# & Ion & Wavelength$^\dagger$ & EW(mod)$^\ddagger$ & EW(obs)$^\ddagger$ &
$v^{\rm obs}_{\rm out}/v^{\rm mod}_{\rm out}$ $^\bigtriangleup$  & EW(obs)/EW(mod)  }
\startdata
1 & \fexxv\ He$\alpha$  & 1.8505 & 5.18 & $3.8\pm1.4$ & 571/2,187 & $0.73\pm0.27$ \\
%\caxix\ He$\alpha$  & 3.1771 & 0.51 & $0.71\pm0.89$ & 290/1236 & $1.4\pm1.76$  \\
2 & \arxviii\ Ly$\alpha$  & 3.7310 & 2.28 & $4.1\pm1.1$ & 192/1,446 & $1.8\pm0.48$ \\
3 & \arxvii\ He$\alpha$  & 3.9493 & 2.94 & $5.0\pm1.1$ & 965/1,009 & $1.69\pm0.37$  \\
%\sxvi\ Ly$\beta$ & 3.9908 & 1.55 & $3.40\pm1.5$ & 1097/1186 & $2.2\pm1.0$ \\
4 & \sxvi\ Ly$\alpha$  & 4.7274 & 12.1 & $10.7\pm1.2$ & 536/1,103 & $0.88\pm0.09$ \\
5 & \sxv\ He$\alpha$  & 5.0387 & 10.9 & $9.2\pm1.2$ & 759/874 & $0.84\pm0.11$ \\
%\sixiv\ Ly$\beta$  & 5.2168 & 5.53 & $9.85\pm2.1$ & 723/850 & $1.78\pm0.38$ \\
%\sixiii\ He$\gamma$ & 5.4046 & 2.74 & $5.65\pm2.9$ & 678/644 & $2.0\pm1.0$ \\
6 & \sixiii\ He$\beta$  & 5.6807 & 5.59 & $6.30\pm1.2$ & 582/670 & $1.12\pm0.21$  \\
7 & \sixiv\ Ly$\alpha$  & 6.1804 & 31.0 & $20.5\pm0.8$ & 755/747 & $0.66\pm0.025$ \\
8 & \sixiii\ He$\alpha$  & 6.6480 & 18.6 & $14.9\pm0.7$ & 798/676 & $0.79\pm0.04$ \\
9 & \mgxii\ Ly$\beta$ & 7.1058 & 7.95 & $8.60\pm0.9$ & 424/624 & $1.1\pm0.11$ \\
10 & \alxiii\ Ly$\alpha$  & 7.1763 & 2.83 & $5.4\pm0.8$ & 883/890 & $1.9\pm0.28$ \\
11 & \mgxii\ Ly$\alpha$  & 8.4192 & 36.6 & $25.1\pm1.4$ & 577/683 & $0.69\pm0.04$ \\
12 & \nex\ Ly$\beta$ & 10.238 & 10.1 & $19.6\pm2.0$ & 664/439 & $1.9\pm0.19$ \\
13 & \nex\ Ly$\alpha$ & $12.132$ &  44.3 & $38.1\pm{2.9}$ & 622/494 & $0.86\pm0.065$ \\
14 & \fexvii\ & 15.014 & 28.2 & $26.2\pm5.3$ & 623/399 & $0.93\pm0.18$ \\
15 & \fexvii\ & 15.262 & 19.1 & $14.9\pm4.3$ & 699/510 & $0.78\pm0.22$ \\
16 & \oviii\ Ly$\alpha$ & 18.969 & 137 & $53.6\pm15.9$ & 925/742 & $0.38\pm0.11$ \\
17 & \ovii\ He$\alpha$ & 21.602 & 74.1 & $40.1\pm34.6$ & 629/333 & $0.54\pm0.46$ \\
18 & \nvii\ Ly$\alpha$ & 24.781 & 96.5 & $72.6\pm34.9$ & 573/387  & $0.75\pm0.36$ \\
\enddata
\vspace{0.05in}
\begin{flushleft}
$^\dagger$ In units of $\aa$. \\
$^\ddagger$ In units of m$\aa$ taken from \citet{K02}. \\
$^\bigtriangleup$ Velocities in units of km~s$^{-1}$.
\end{flushleft}
\label{tab:tab2}
\end{deluxetable}

Besides the observable quantities discussed  so far (as listed in {\bf Table~\ref{tab:tab2}}), we also calculate the
characteristics associated with specific lines of our  magnetically-driven disk-winds.
Representative wind variables are listed in {\bf Table~\ref{tab:tab3}} in order of
increasing characteristic LoS distance, $r_c$, (second column) where the local ionic
column obtains its maximum value for each line.
We also show the local wind density $n_c$ (fifth column)  and line optical depth
$\tau_c$ (seventh column) at this characteristic radius $r_c$.  Accordingly, we keep
track of the radial dynamic range over which the local ionic column stays within $50\%$
of its maximum value. Within this range, the ionization parameter $\xi$ (third column)
and the wind temperature $T$ (fourth column) are computed as well. Lastly, the LoS-integrated hydrogen equivalent column density of each ion $N_H^{\rm tot}$ is also calculated (sixth column). As
shown, the distances of absorbers range from 0.2 pc to 30 pc scales in X-rays, exactly
as estimated for this source by \citet{Reeves04}, \cite{Behar03} and \cite{Gabel05}. In
most of the radial extent of the wind, absorbers are optically thin where the radiative
transfer calculations with {\tt xstar} are justified.

It should be reminded that continuous outflows are characterized, in general, by a
gradient of the wind's physical quantities (e.g.density, velocity etc.), a situation
fundamentally distinct from outflow models consisting of distinct, independent,
multiple kinematic and ionization components. In the continuous outflow situation all
wind elements leave their imprint on the line profile at their respective velocities,
weighted of course by the corresponding ionic abundance. For example, in the specific
wind structure considered here, the ionic abundances are not symmetric with respect to
the value of $\xi$ where they achieve their maximum: they fall much sharper at lower
$\xi$-values (and lower velocities) than at higher ones (higher velocities) (see one of the template AMDs as shown in {\bf Fig.~1}). As a result, the resulting absorption lines are
asymmetric, skewed toward their blue side and their peak absorption is at a velocity
larger than that corresponding to their peak column (see \S 4.1 below).

\begin{figure}[h]% ------------------------------------- Figure~6
\begin{center}$
\begin{array}{c}
%\includegraphics[trim=0in 0in 0in
%0in,keepaspectratio=false,width=3.4in,angle=-0,clip=false]{chi2_2.pdf}
\includegraphics[trim=0in 0in 0in
0in,keepaspectratio=false,width=3.1in,angle=-0,clip=false]{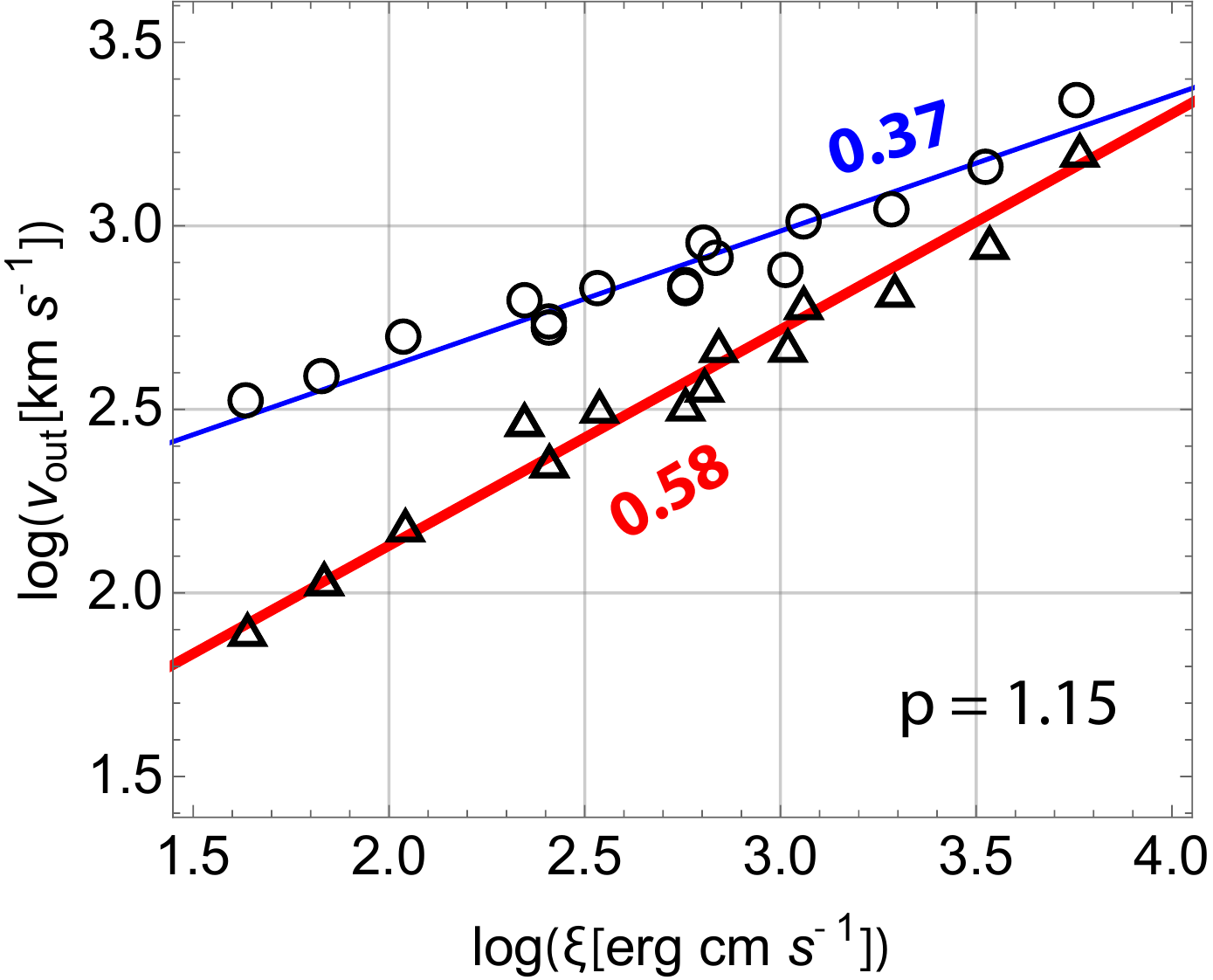}\includegraphics[trim=0in 0in 0in
0in,keepaspectratio=false,width=3.2in,angle=-0,clip=false]{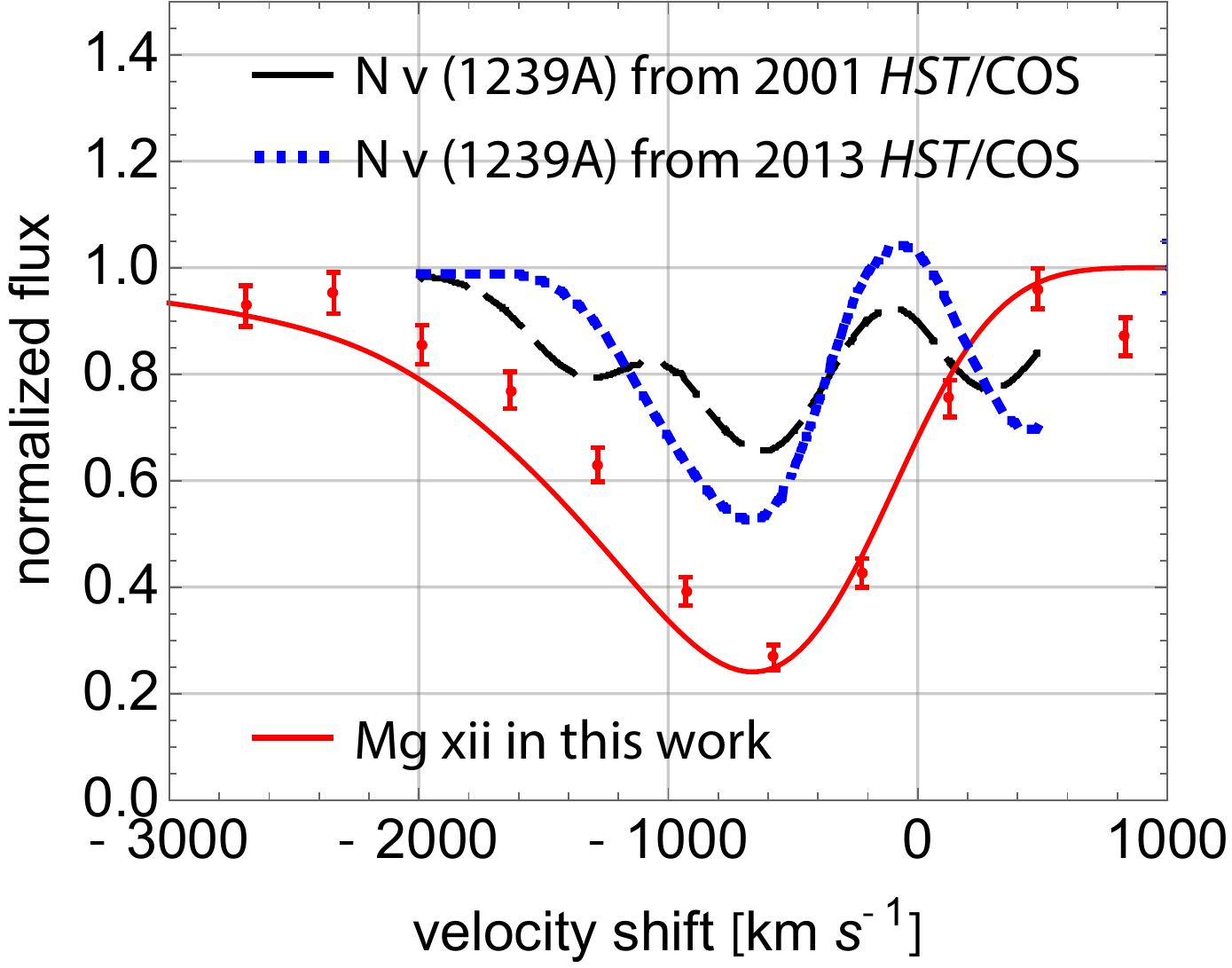}
\end{array}$
\end{center}
\caption{(a) The observed correlation between the LoS outflow speed
$v_{\rm out}$ [km~s$^{-1}$] and ionization parameter $\xi$ [erg~cm~s$^{-1}$] for the
same 18 ions in {\bf Tables~\ref{tab:tab2}-\ref{tab:tab3}}. Triangles denote a correlation with the velocity at which the ionic column derived from our model is locally maximum for each ion, while circles show a correlation for the
same ions when the velocity is simply determined by the trough of the bestfit spectrum
model for each ion in {\bf Figures~\ref{fig:spec1}-\ref{fig:spec2}}. The blue line denotes the corresponding linear regression of slope $0.37$ for the circles.  Note that the predicted scaling (in red) from equation~(\ref{eq:vel}), $v_{\rm out} \propto \xi^{1/\{2(2-p)\}} = \xi^{0.58}$ for $p=1.15$, agrees
very well with the bestfit correlation (triangles). (b) The solid red curve shows the normalized velocity profile of \mgxii\ Ly$\alpha$
($8.4192\aa$; red) of the bestfit model $p=1.15$ wind of this work. The dashed (dark) and dotted (blue) curves respectively represent the profiles of the UV \nv\ absorber ($1239\aa$) from {\it HST}/COS 2001 and 2013 observations in \citet{Scott14}. The UV spectra are intentionally smoothed to the velocity
resolution ($\sim 100$ km~s$^{-1}$) of {\it Chandra}/HETG for comparison. }
\label{fig:cor}
\end{figure}

%------------------------------- Table~3
\begin{deluxetable}{lc|cccccc}
\tabletypesize{\small} \tablecaption{Bestfit Photoionization Modeling for 18 Absorbers
}
%\tablewidth{0pt}
\tablehead{\# & Ion & $\log(r_c/r_S)^\dagger$ & $\Delta (\log \xi)^\ddagger$ & $\Delta
(\log T)^\Diamond$ & $\log n_c$$^\sharp$  & $N_H^{\rm tot}$$^\bigtriangleup$ & $\tau_c$}% \\
%& & & & & & [$10^{21}$ cm$^{-2}$] & [$10^{-3}$] }
\startdata
%
%1 & \nixxviii\ Ly$\alpha$  & 4.24 & 3.9 - 5.0 & 6.2 - 6.6 & 6.2 & 12.7 & 1.8 \\
%2 & \fexxvi\ Ly$\alpha$    & 4.42 & 3.7 - 4.9 & 6.1 - 6.6 & 6.7 & 12.2 & 62 \\
%3 & \fexxvi\ Ly$\beta$     & 4.42 & 3.7 - 4.9 & 6.1 - 6.6 & 6.7 & 12.5 & 10 \\
%4 & \coxxvii\ Ly$\alpha$   & 4.42 & 3.8 - 4.9 & 6.2 - 6.6 & 6.7 & 11.9 & 0.30 \\
%5 & \mnxxv\ Ly$\alpha$     & 4.60 & 3.7 - 4.7 & 6.1 - 6.6 & 6.5 & 10.9 & 0.83 \\
%6 & \mnxxv\ Ly$\beta$      & 4.60 & 3.7 - 4.7 & 6.1 - 6.6 & 6.5 & 10.9 & 0.13 \\
%7 & \crxxiv\ Ly$\alpha$    & 4.70 & 3.6 - 4.6 & 6.1 - 6.5 & 6.4 & 10.4 & 2.7 \\
%8 & \crxxiv\ Ly$\beta$     & 4.70 & 3.6 - 4.6 & 6.1 - 6.5 & 6.4 & 10.4 & 0.43 \\
%9 & \coxxvi\ He$\alpha$    & 4.88 & 3.8 - 4.2 & 6.2 - 6.4 & 6.2 &  5.9 & 1.1 \\
%\nixxvii\ He$\alpha$  & 4.78 & 3.5 - 4.3 & 6.0 - 6.5 & 5.6 & 7.5  & 0.39 \\
1 & \fexxv\ He$\alpha$     & 4.96 & 3.5 - 4.2 & 6.0 - 6.4 & 6.1 &  6.1 & 0.27 \\
%11 & \caxx\ Ly$\alpha$      & 4.96 & 3.3 - 4.3 & 6.0 - 6.4 & 6.1 & 10.0 & 14 \\
%12 & \caxx\ Ly$\beta$       & 4.96 & 3.3 - 4.3 & 6.0 - 6.4 & 6.1 & 10.0 & 2.3 \\
%13 & \mnxxiv\ He$\alpha$    & 5.06 & 3.4 - 4.1 & 6.0 - 6.3 & 6.0 & 5.6  & 3.1 \\
%14 & \crxxiii\ He$\alpha$   & 5.14 & 3.3 - 4.0 & 6.0 - 6.4 & 5.9 & 5.6 & 11 \\
%15 & \fexxiv\ Li$\alpha$    & 5.14 & 3.3 - 3.9 & 6.0 - 6.2 & 5.9 & 1.8 & 26 \\
%\fexxiv\ Li$\alpha$  & 5.7 & 3.7 - 3.9 & 6.1 - 6.2 & 5.2 & 2.3 & 17.0 \\
2 & \arxviii\ Ly$\alpha$   & 5.2  & 3.0 - 4.1 & 5.9 - 6.3 & 5.8 & 9.5 & 0.048 \\
3 & \arxvii\ He$\alpha$    & 6.7  & 2.7 - 3.5 & 5.8 - 6.0 & 5.4 & 5.7 & 0.22 \\
%17 & \arxviii\ Ly$\beta$    & 5.2  & 3.0 - 4.1 & 5.9 - 6.3 & 5.8 & 9.5 & 7.7 \\
%18 & \caxix\ He$\alpha$     & 5.5  & 3.0 - 3.7 & 5.9 - 6.1 & 5.5 & 5.5 & 62 \\
4 & \sxvi\ Ly$\alpha$      & 5.5  & 2.8 - 3.9 & 5.8 - 6.2 & 5.5 & 8.7 & 0.27 \\
5 & \sxv\ He$\alpha$       & 6.0  & 2.5 - 3.3 & 5.7 - 5.9 & 4.9 & 4.6 & 0.12 \\
6  & \sixiii\ He$\beta$    & 6.3  & 2.2 - 2.9 & 5.6 - 5.8 & 4.5 & 4.1 & 0.76  \\
7 & \sixiv\ Ly$\alpha$     & 5.8  & 2.5 - 3.7 & 5.7 - 6.1 & 5.1 & 8.1 & 1.0 \\
8 & \sixiii\ He$\alpha$    & 6.4  & 2.2 - 3.0 & 5.6 - 5.9 & 4.5 & 4.0 & 4.4 \\
9 & \mgxii\ Ly$\beta$      & 6.1  & 2.3 - 3.4 & 5.6 - 6.0 & 4.8 & 4.8 & 0.27 \\
10 & \alxiii\ Ly$\alpha$    & 6.1  & 2.3 - 3.4 & 5.6 - 6.0 & 4.8 & 6.7 & 0.05 \\
11 & \mgxii\ Ly$\alpha$     & 6.1  & 2.3 - 3.4 & 5.6 - 6.0 & 4.8 & 4.8 & 1.7 \\
12 & \nex\ Ly$\beta$        & 6.5 & 2.0 - 3.0  & 5.0 - 5.8 & 4.3 & 4.9 & 0.38 \\
13 & \nex\ Ly$\alpha$       & 6.5 & 2.0 - 3.0  & 5.0 - 5.8 & 4.3 & 4.9 & 2.4 \\
14 & \fexvii\               & 6.6  & 2.1 - 2.6 & 5.5 - 5.7 & 4.2 & 1.3 & 18 \\
15  & \fexvii\              & 6.6  & 2.1 - 2.6 & 5.5 - 5.7 & 4.2 & 1.3 & 4.5  \\
16  & \oviii\ Ly$\alpha$    & 6.9  & 1.6 - 2.6 & 4.6 - 5.7 & 3.8 & 4.3 & 100 \\
17  & \ovii\ He$\alpha$     & 7.4  & 1.2 - 1.6 & 4.4 - 4.8 & 3.3 & 3.9 & 547 \\
18  & \nvii\ Ly$\alpha$     & 7.1  & 1.4 - 2.2 & 4.5 - 5.6 & 3.6 & 4.3 & 13 \\
%
%20 & \sxvi\ Ly$\beta$       & 5.5  & 2.8 - 3.9 & 5.8 - 6.2 & 5.5 & 8.7 & 42 \\
%
%23 & \sixiv\ Ly$\beta$      & 5.8  & 2.5 - 3.7 & 5.7 - 6.1 & 5.1 & 8.7 & 160 \\
%
%
%
\enddata
\vspace{0.0in}
\begin{flushleft}
$^\dagger$ Characteristic LoS distance $r_c$ where the modeled AMD of
specific charge
state becomes maximum (i.e. $\Delta N_H/\Delta (\log \xi)$ is maximum). See the text for details. \\
$^\ddagger$ Range of ionization parameter $\xi$ [erg~cm~s$^{-1}$] over which the local
ionic column per ionization parameter bin is greater than $50\%$ of the locally maximum ionic column. \\
$^\Diamond$ Range of temperature $T$ [K] over which the local ionic column per ionization
parameter bin is greater than $50\%$ of the locally maximum ionic column. \\
$^\sharp$ Characteristic number density $n_c$ [cm$^{-3}$] of ion at $r=r_c$. \\
$^\bigtriangleup$ LoS-integrated column $N^{\rm tot}_H$ [$\times 10^{21}$cm$^{-2}$] of ion . \\
%$^\ast$ Characteristic line optical depth $\tau_p$ [$\times 10^{-3}$] of ion. \\
\end{flushleft}
\label{tab:tab3}
\end{deluxetable}

\subsection{The $v_{\rm out} - \xi$  Correlation }

The continuous variation of $\xi$ and outflow velocity $v_{\rm out}$ with distance $r$ implies a
correlation between $\xi$ and the projected outflow velocity along the observer's LoS
$v_{\rm out}$ (i.e. eqn.~(\ref{eq:vel})). Such a relation has been searched empirically in the data.
For example, \citet[][]{Detmers11} found the correlation $v_{\rm out} \propto
\xi^{0.64\pm0.1}$ in the data of the 600-ks {\it XMM-Newton}/RGS multiwavelength
campaign of Mrk~509 roughly consistent with our scaling value in Eq. (3) for $p=1.15$.
\citet{Tombesi13} has conducted a systematic analysis on the  observed WAs and UFOs
from a sample of 35 Seyfert 1 galaxies and found that $v_{\rm out} \propto \xi^{0.65}$
for the WAs and UFOs combined together, while $v_{\rm out} \propto \xi^{0.31}$ for the
WAs only (see also \citealt{Laha14} for a similar analysis).
However, before strong conclusions are drawn, one should first bear in mind that all
the points shown in \citet{Tombesi13} represent data from a large number of AGN of
different inclinations, columns and SEDs. The derived scaling law, equation~(\ref{eq:vel}), appears to
be generic, but it should be reminded (as mentioned elsewhere in this paper) that the
magnetic field and velocity structure of our model winds  are self-similar, laterally
stacked paraboloids, while the ionizing radiation (at least in our calculations so far)
is spherically symmetric about the AGN. Therefore, deviations from the generic relation
of equation~(\ref{eq:vel}) should not be surprising in a more detailed modeling. %\st{This is precisely
%what is depicted in Figure~4b: }
%This is precisely what is depicted in {\bf Figure~\ref{fig:f4}b}:
%

{\bf Figure~\ref{fig:cor}a} depicts such $v_{\rm out} - \xi$ correlations based on two
different standpoints.
The triangles represent the correlation of ion velocity $v_{\rm out}$ with the value of
$\xi$ at a characteristic radius, $r_c$ (see {\bf Table~3}), where each ion achieves its
maximum column density by photoionization along the LoS  (see {\bf Fig.~\ref{fig:amd}});
these points are indeed in excellent agreement with the expected relation of equation~(\ref{eq:vel}) of slope of $0.58$ (depicted by red line) for the bestfit $p=1.15$ wind.
However, the line optical depth $\tau_{\rm ion}$ may not necessarily be the greatest at
$r=r_c$, as the photo-absorption cross section $\sigma_{\rm abs}$ given by equation~(\ref{eq:sigma}) is
monotonically increasing with distance causing the peak of the line depth to shift
slightly outwards along the LoS. Hence, the outflow  velocity derived from the model
spectral line trough is systematically lower than the velocity at $r=r_c$; however this
shift cannot be deduced without detailed line modeling.
%
%Thus, it is customary in general to use a centroid wavelength to refer to the wind velocity.
Thus, as is customary, we use a centroid wavelength to refer to the wind LoS velocity.
The results of these measurements based on the line troughs are represented by the circles in Figure~\ref{fig:cor}a; as already seen in Figures 2 and 3, these data agree well with the expected velocities from the best-fit model. Their velocities appear to be well
correlated to the ionization parameter $\xi$, but its slope is different from
the simple analytic prediction of $v_{\rm out} \propto \xi^{1/\{2(2-p)\}}$.
%
%The results of such modeling are represented by the circles in the figure, and
%correspond to velocity associated with the troughs of the lines of the best-fit model
%(shown in {\bf Figs.~\ref{fig:spec1} and \ref{fig:spec2}}) for each ion and are in good
%agreement with the data.
%
The best-fit linear regression yields $v_{\rm out} \propto
\xi^{0.37}$ (denoted by blue line). This is in fact very close to that from the WA ($\simeq 0.31$)
quoted in \citet{Tombesi13}. One cannot fail to notice that this is much flatter than
the one would  naively expect from equation~(\ref{eq:vel}).
%
%Hence, we find in this analysis that such a seemingly simple  correlation analysis
%could be misleading.

In summary, one should note that detailed modeling of the line profiles of individual
objects are necessary, along the lines of a well defined model, in order to decide on
the implications of the $v_{\rm out} - \xi$ correlations of specific observations. We
defer general statements on this issue until we have modeled this correlation for a
wide range of wind parameters, the subject of a future publication. Finally, one should
bear in mind that our models represent a self-similar \emph{steady-state} model. We do
know that AGN are variable, both on short and long time scales. Our models assume that
variations in the disk mass flux are also reflected in mass flux of the wind; these
variations are likely to break the simple power law density profiles assumed so far,
producing, for example, increased absorption in earlier less absorbed spectra \citep[e.g.][]{Mehdipour17}. Also, factors other than magnetic fields may influence the wind
mass flux; for example, increased mass flux beyond that of self similarity at the inner
wind regions due, e.g. to radiation pressure, would produce higher columns for the
highest $\xi$ ions, in agreement with the \citet{Tombesi13} compilation.

\clearpage

\section{Discussion \& Conclusions}

{The 900-ks {\it Chandra}/HETG spectrum  of the luminous, radio-quiet Seyfert 1 AGN,
\obj, is analyzed with emphasis on its X-ray absorption spectrum which was modeled
within the framework of magnetically-driven accretion-disk winds. With its AMD already
determined in B09, in conjunction with earlier extensive spectral studies (e.g. K01,
K02 and \citealt{Krongold03}), we focused on the $\sim 2\aa - 20\aa$ X-ray band to
model the properties of its most important transitions in this range. Comparison of the
absorber properties to observation, then, provides the global parameters of this wind,
namely its density index $p$, its normalization $n_o$ at its innermost launch radius
(of order $r_S$) and its inclination angle $\theta_{\rm obs}$. Thus we found, in the context
of this model, that the wind radial density profile is given by $n(r) \simeq 6.9 \times
10^{11} ~ (r/r_S)^{-1.15}$ cm$^{-3}$, along with $\theta \simeq 44\deg$ for the
inclination angle. One should note that the value of the density profile index $p =
1.15$ is consistent with that inferred in B09 obtained on the basis of the apparent
absorption line column $N_H$. However, the present analysis includes additional
kinematic information (see {\bf Figs.~\ref{fig:spec1}-\ref{fig:spec2}})  via mutually coupled ions that affects the shape and EW of the observed
lines, features that allow for a refinement of the precise value for $p$.
The obtained values of $(p,n_{11})$ as well as the inclination angle $\theta_{\rm obs}$ are determined by fitting the spectroscopic observations assuming a radially self-similar wind model. These constraints are therefore consistent and reasonable with the spectral results.

This is our second application of the specific MHD wind formalism to determine
their global properties through modeling the multi-line absorption X-ray spectra of
accreting black holes, our previous analysis being that of the spectrum of the
galactic X-ray black hole binary GRO~J1655-40 \citep{F17}. We have argued elsewhere
\citep[e.g.][]{K12} that these models are scale-free and should be applicable to any accreting
black holes. We are very encouraged by the ability of this most simple, self-similar,
model to reproduce the properties of absorption features of outflows associated
with objects of such disparate mass scales. In this section we discuss the results and
conclusions of our present analysis and their relation to similar observations of this
and other AGNs.

In fact, it has been long speculated that some Seyfert AGNs may exhibit ionized winds whose physical characteristics indeed favors for MHD-driven scenario. For example, \cite{Turner05} found a series of X-ray absorption lines in {\it Chandra} grating data and investigated the nature of their physical properties. They argue that some absorbers are too highly ionized to be radiatively accelerated, which can be a circumstantial evidence for a hydromagnetic origin for the outflow in NGC~3516.
From a synergistic analysis of another well-known Seyfert 1 galaxy, NGC~4151, by \cite{Couto16}, some components of the observed X-ray absorbers are highly ionized. Given the observed 2-10 keV flux and photoionization modeling, the calculated force multiplier  is found to be too small to drive the observed winds, perhaps indicating a magnetic-origin. These observations may be pointing to the relevance of MHD-driven process at least in part if not fully, as discussed in this work for \obj\ as well as our earlier work for the XRB, GRO~J1655-40.

It should be noted that our model treats the underlying accretion disk as a boundary condition that provides the seed plasma for the winds. Hence, accretion and outflows are solved independently. In reality, however, the inflow-outflow problem must be self-consistently considered. Some authors \citep[e.g.][]{Ferreira97,Casse00a,Casse00b,Chakravorty16} have attempted this problem by assuming that magnetic flux is brought in from infinity and its advection is balanced by its diffusion. They show that this requirement tends to lead to a very steep density profile of $p \simeq 3/2$ inconsistent with the X-ray AMD observations (e.g. B09), as discussed in this paper. Under a certain ionization parameter space, on the other hand, they argue that a less steep profile ($p \sim 1.1$), as favored in our work here, could be obtained. The qualitative argument for winds such as considered here is that the viscous torques that transfer outward the disk angular momentum transfer also mechanical energy (and possibly magnetic flux; see \citealt{Contopoulos17}). It is this energy that powers the wind mass flux.
Although the mutual coupling among inflows, outflows and threaded magnetic fields is fundamentally important, it is beyond the scope of our current work.

Among the X-ray AGN absorption  features, \fexxv\ and  \fexxvi\ have attracted
particular attention because of their ubiquity and high velocities as in the UFOs \citep{Tombesi13}.
These properties have prompted their study in \obj\ also by \emph{XMM-Newton}
\citep{Reeves04}. 
Fits of the broader \fexxv\ band with multi-component, photoionized
plasma provide absorber parameters similar to ours. However, in similarity with Mrk~509, the \fexxv\ feature is weak in this AGN too. We speculate that this is the result
of the relatively low inclination angle and the SEDs in both objects.
While systematically consistent for many ions, the observed velocities of high-Z
ions, such as Fe and Ar, appear to be lower than what is predicted by the model (i.e.
$\sim 1,000-2,000$ km~s$^{-1}$; see Table~2 and Fig.~3). As the characteristic of the
wind model, heavier absorbers (e.g. Fe) emerge at smaller distances where the
velocities are higher. As a consequence, the resulting absorption features are exected to be
broader. 
There are several ways to account for the discrepancy that may involve all of them; First of all, the error bars of these line transitions are much larger than those of longer wavelengths making this discrepancy less significant. Secondly, our calculations have ignored scattering which may fill in some of the line profiles (note that this would occur preferentially at the shorter wavelengths of the profiles as the scattering depth is larger at smaller distances and the scattering cross section competes with that of photoabsorption). Finally, break-down of self-similarity near the axis at small radii and/or of the smoothness of the wind geometry may also contribute to this discrepancy.
While beyond the scope of the current paper, we plan to address this issue quantitatively in a future work.

%We speculate that the scattered continuum and/or line photons in this denser
%part of the wind (coming from outside of the LoS) might fill in the otherwise broader
%trough feature making it seemingly narrow as detected. 

An important set of observations additional to those discussed herein have
been those of \cite{Mehdipour17}, who analyzed the multiwavelength spectra of \obj\
during a recent, reduced soft X-ray ($E \lsim 1$ keV) flux state in order to study the
properties and physics of obscuring plasma in this AGN. They found that obscuration of
the X-rays at $E \lsim 1$ keV produced also more prominent \fexxvi\ and \fexxv\
absorption. They interpreted their results as due to the effects of an intervening,
highly ionized obscurer. We defer detailed analysis of this latest data to a future
publication. However, we would like to point out that within our framework there may be
a natural explanation, considering that increased absorption column is due to increased
\emph{poloidal mass flux}. Such a view is supported by the fact that the optical and UV
(O-UV) continua of their 2016 observations are a factor of $\simeq 2$ higher than that
of the early 2000's ones (the so-called ``unobscured state") analyzed in the present
paper (while their hard X-ray fluxes have so far remained unchanged). If the wind and
disk mass fluxes vary in unison, as implicit in our model, the excess mass flux of the
absorber (i.e. the wind) is then related to increased mass flux in the disk that drives the
O-UV emission of this object. If the O-UV spectrum is emitted at radii larger than
those of the X-rays \citep{Chartas09b,K15} and the excess mass is slowly being accreted
inward, one should expect an increase in the X-ray flux to follow that of the O-UV,
with corresponding increase in soft X-ray ionization.

\subsection{Mass-Accretion for Disk-Winds}

Our wind model predicts that the wind mass flux scales with distance as
$\dot{M}(x) \propto x^{3/2-p}$ where $x \equiv r_/r_o \sim r/r_S$. This means that, for
$p<3/2$, the wind mass flux increases with distance; i.e. most of the mass available
for accretion is lost into the wind at large distances from the black hole. This
implies that the disk accretion rate should decrease toward the black hole. While our
models do not address this problem, they have to allow for accretion onto the black hole
at a rate sufficient to produce the observed bolometric luminosity. To provide an
approximate resolution of this issue, we assume that the ratio of the wind mass flux to that
of the disk is unity, $f_w \simeq 1$, independent of the radius \citep[see F10a;][]{F17}. In this way, we
can connect the mass flux at the outer edge of the disk\footnote[1]{This part of the wind may be viewed as the putative torus in the context of a unified torus scenario \citep[e.g.][]{KK94,K12}.}  to that producing
the observed luminosity through accretion onto the black hole. This argument then
implies
\begin{eqnarray}
\dot{M}_{\rm}(x=x_{\rm max} \simeq 10^6) \sim 10^{3(3-2p)} \dot{M}(x=1) \ , \label{eq:mdot}
\end{eqnarray}
where $\dot{M}(x=1) \simeq n_o \sigma_T r_S \dot{M}_E$ is given in terms of the wind
density normalization $n_o$ with $\dot{M}_E \equiv L_E/c^2$ being the Eddington mass-flux
rate, $L_E$ being the Eddington luminosity and $\sigma_T$ being the Thomson cross section. For our
bestfit model with $p = 1.15$ and $n_{11} = 6.9$ and for a black hole mass $M=3 \times
10^{7} \Msun$, one obtains
\begin{eqnarray}
\dot{M}(x \simeq 1) &\sim&  4 \dot M_E \simeq 1.8 \times 10^{27} ~~ \textmd{g~s$^{-1}$}  \ , 
\\
\dot{M}(x_{\rm max} \simeq 10^6) &\sim& 500 \dot M_E \simeq 2.2 \times 10^{29}
~~ \textmd{g~s$^{-1}$}  \ ,
\end{eqnarray}
where $\dot{M}(x_{\rm max} \simeq 10^6)$ is available over an extended region ($x \lesssim 10^6$) of the disk to be
provided for winds. 
Liberated accretion power via gravitational potential energy, given by $L \propto
GM \dot{M} / r \propto x^{1/2-p}$, is more significant at {\it smaller} distances contrary to the dominant mass-accretion rate $\dot{M}_{\rm}$ at {\it larger} radii in equation~(\ref{eq:mdot}).  
Similarly, wind kinetic luminosity scales as 
\begin{eqnarray}
L_{\rm wind} \sim \dot{M} v_{\rm out}^2 \propto x^{1/2-p} \ ,
\end{eqnarray}
and hence this power is very small ($\simeq 10^{-4}$) at $x_{\rm max} \simeq 10^6$ for $p=1.15$, despite its
large mass flux there, compared to that carried onto the black hole at $x \sim 1$.
This fact has also been noted by \cite{KraemerCrenshaw12} which, using
UV photonionization considerations for 7 Seyfert galaxies including \obj, estimated that the UV absorber's mass flux is $\sim
100$ times larger than that is needed to power the observed luminosity by accretion onto
the black hole in consistence with our finding. 

The bestfit value of $n_{11}=6.9$ we have obtained implies Thomson depth of order
$\gsim 1$ at the wind's innermost radius. For $f_w \simeq 1$, the corresponding  accretion
kinetic luminosity is $\sim 10^{46}$ erg~s$^{-1}$ (proportionally smaller for a smaller black
hole mass), and considering the efficiency of a \sw black hole ($\eta \simeq 0.05$, and
maybe a little smaller if radiation is trapped in the flow), we obtain a bolometric
luminosity of $\sim 5 \times 10^{44}$ erg~s$^{-1}$; considering its apportionment across the
entire electromagnetic spectrum, the ionizing luminosity of $L_X \sim 3 \times 10^{43}$ erg~s$^{-1}$ employed in our modeling seems to be in a reasonable agreement with the model's global mass flux budget and observations \citep[see, e.g., Fig.~6 in][]{Mehdipour17}.

\subsection{Physical Link: UV/X-ray Absorbers and NLR Outflows}

One of the {open issues} regarding AGN winds is {the} link between their X-ray
absorbers with the known UV absorbers (e.g. \civ\ and \nv) found in {\it HST}/COS/STIS
observations \citep[e.g.][]{Crenshaw99,CKG03,CrenshawKraemer12}. Despite a number of
analyses supporting this relation to date \citep{Mathur94, Mathur95, Crenshaw99,Collinge01, Kraemer02,CKG03, Krongold03, Kaastra14}, an explicit physical description  of the underlying plasma dynamics is missing.

The simultaneous presence of {\it HST, Chandra} and {\it XMM-Newton} in orbit has
provided the opportunity of the synergistic study of absorbers in the X-ray and UV
regions of the spectra of several AGNs. For example, Collinge et al.~(2001) showed that
the lower ionization, X-ray Fe absorption features of NGC 4051 had corresponding UV
counterparts, while the higher-ionization, higher-velocity X-ray absorbers of the
spectrum lacked an equivalent UV absorption, indicating the absence of these ions in
the higher ionization, higher velocity plasma. With respect to \obj, \citet{Gabel03}
found the UV absorption features to have velocity structures similar to their
X-ray counterparts, thereby arguing for the continuity of both absorption components.

In {\bf Figure~\ref{fig:cor}b}, we show the data and our model profile of the \mgxii\ line
(solid red), overlaid on the observed absorption lines of \nv\ (dashed dark for 2001 data and dotted blue for 2013 data) obtained by
\emph{HST}/COS data (from \citealt{Scott14}; see their Fig.~9) smoothed to the {\it
Chandra}/HETG resolution ($\sim 100$ km~s$^{-1}$) for fair comparison. 
It is interesting that both UV and X-ray lines exhibit approximately the {\it same velocity
structure} (i.e. trough position and width) and the {\it same absorption depth}
($\tau_{\rm \nv}\simeq 0.45, \; \tau_{\rm \mgxii}\simeq 1.5$) considering the difference in ionic columns produced.
It is noted that 2001 UV \nv\ line appears to exhibit a multiple trough feature. As discussed in \cite{Scott14}, it is conceivable that poorer spectral (velocity) resolution of X-ray measurements with {\it Chandra}/HETG may limit a detection of finer (smaller) kinematic components of X-ray absorbers. Nonetheless, the overall kinematic component seen in both UV (\nv) and X-ray (\mgxii) observations looks surprisingly similar in depth and broadness in general. Since our model is focused primarily on X-ray winds, we will not discuss the UV absorbers any further.
This would generally be
an unlikely situation, because the two lines ``live" in very different regions in
$\xi$-space (of $\log \xi \simeq 1$ for \nv\ and $\log \xi \simeq 2-3$ for \mgxii, respectively) and the
\nv\ has much higher absorption cross section. One could consider that their similar
velocity structure argues for these ions belonging to the asymptotic velocity (and also
asymptotic $\xi$) region of a spherically symmetric wind whose mass flux can be
arranged so that the ionic abundances of these two ions are roughly inversely
proportional to their cross sections; this would produce similar line kinematics and
depths. However, such a wind would preclude the presence of higher-$\xi$,
higher-$v_{\rm out}$ ions such as \fexxvi.

On the other hand, a UV/X-ray absorption line similarity is possible within our model
if we assume that the wind terminates at a distance $r$, or equivalently, at an
ionization parameter $\xi$, where \nv\ is still subdominant such that the ion
abundances of these two ions are roughly inversely proportional to their cross section
(since the \nv\ column is smaller than that of \mgxii, these are only broad estimates);
then, their similar profiles will reflect the kinematic properties of this zone while
maintaining roughly similar depths.

We have extended our  X-ray ionization calculations to the UV ionization zone of \nv\ to find a line
absorption depth $\tau_{\rm \nv} \simeq 0.35$ at $r/r_S \sim 10^6$ and $v_{\rm out}
\sim 500 \, {\rm km \; s}^{-1}$. The termination of the wind takes place in our model
at a distance of $\sim 1$ pc along the LoS. An extension of the (self-similar) wind to
larger distances would result in much lower \nv\ velocity and high absorption \nv\
depth that are not observed. Finally, one additional consideration in comparing the
O-UV and X-ray absorption line profiles is that the UV source region is likely to be
larger than that of the X-rays \citep{Chartas09b}; therefore their profiles do not
necessarily correspond to the same LoS and velocity structures. Also, the UV source may
not be totally covered by the wind if it is clumpy.

\begin{figure}[ht]% ------------------------------------- Figure~7
\begin{center}$
\begin{array}{c}
\includegraphics[trim=0in 0in 0in
0in,keepaspectratio=false,width=3.35in,angle=-0,clip=false]{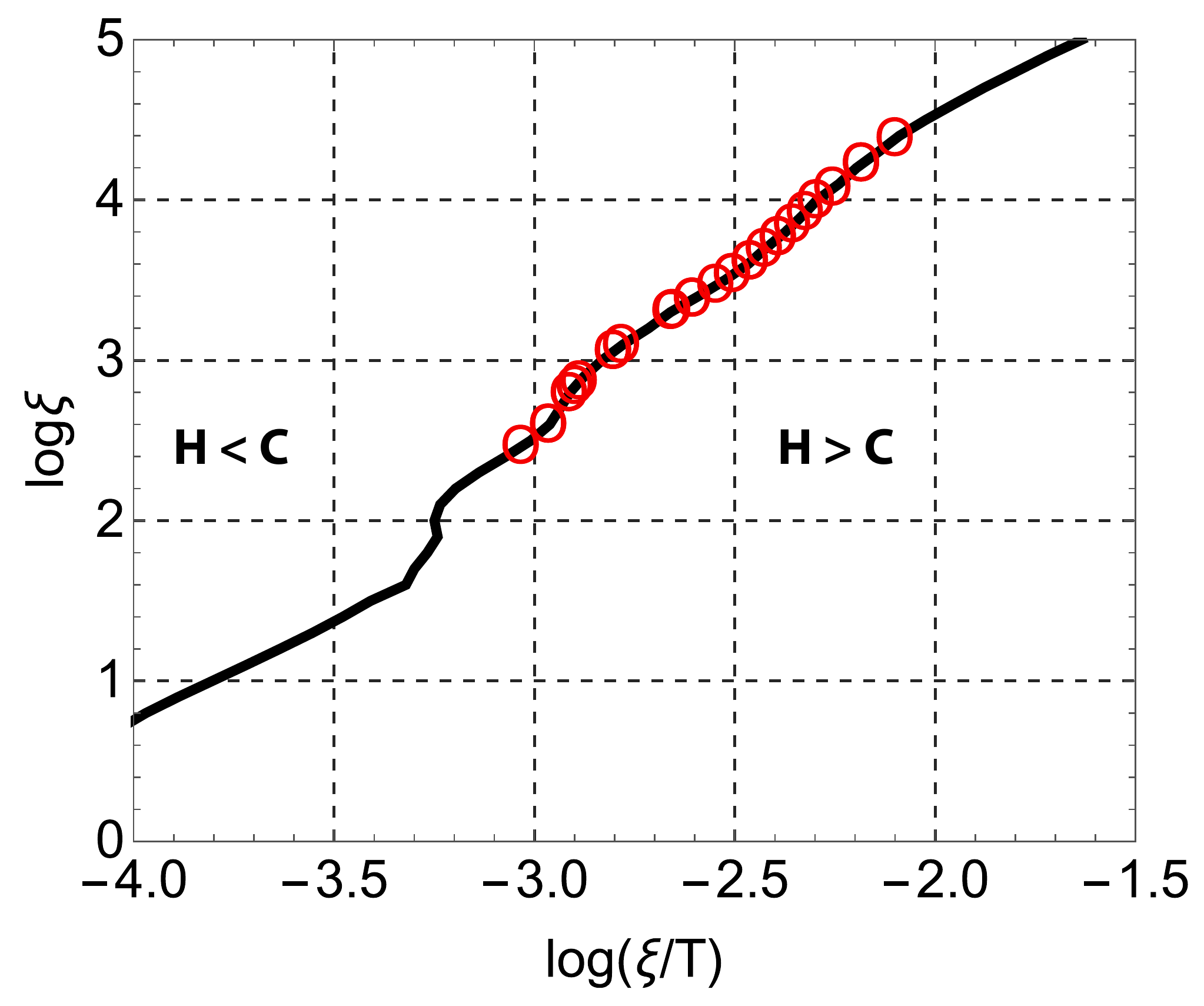}
\end{array}$
\end{center}
\caption{Modeled cooling curve (in black) derived from the photoionization calculation
with the bestfit wind solution of $p=1.15$. Circles (in red) indicate the positions of 18
absorbers listed in {\bf Tables~2-3} in the parameter space. Regions where heating (H) or cooling
(C) dominates are also indicated. } \label{fig:cooling}
\end{figure}

%[patchy clouds models:]

\subsection{Clumpy Absorbers}

An altogether different scenario to produce the observed AMD distribution is that of
AGN clouds (clouds have been a basic staple of AGN phenomenology, especially since the
work of \citealt{Krolik81}) that are not uniform but include a density stratification
\citep[e.g.][]{Goncalves06,Rozanska06,Adhikari15,Stern14,Goosmann16}. This alternative
view is mostly prompted by lacking of some ions in the observed AMD over a finite
narrow ionization parameter space; e.g. at $\log \sim 1 - 2$ for IRAS~13349+2438 and
\obj\ \citep[e.g.][]{HBK07} while $\log \xi \sim 2 - 2.8$ for Mrk~509
\citep[e.g.][]{Detmers11}. Since clouds are thought to be the products of a thermal
instability occurring in X-ray photoionized plasmas under constant pressure, the
absence of transitions at specific $\xi$-values is often attributed to this instability.
While these considerations have not been incorporated in our calculations, we would
like to mention that our winds operate under conditions of constant density (locally),
a situation not conducive to the aforementioned instability \citep[e.g.][]{Rozanska06}.

{\bf Figure~\ref{fig:cooling}} shows a calculated cooling curve for the bestfit photoionized
wind of $p=1.15$ under constant density thermal equilibrium; no thermal instability is
apparent, typically indicated by a double-valued {\tt S-shaped} region  in the
$\xi-\Xi$ space where $\Xi \equiv \xi/T$. The positions of the  ions considered in this work (see {\bf Tables~\ref{tab:tab2}-\ref{tab:tab3}}) are marked by the circles on the cooling curve at the
derived values of $\xi$ and $T$. Generally, under the thermal instability conditions,
the observed ions should be associated with clouds in pressure equilibrium with hot
gas. However, these occur under rather narrow regions in $\xi$ inconsistent with the
broad range of ionization values inferred from the observed transitions.
\section{Summary}

The line profiles of {\bf Figures~2-3} highlight the success of our theoretical approach to model MHD-driven absorbers: (1) The
properties of individual lines (i.e. shape and EW) are not independent; they were all
computed with the same (and also help define the) global parameters of a wind that
spans $10^5-10^6$ \sw radii in space and hence their physical conditions are all
mutually coupled. (2) The detailed profiles of the lines are not symmetric (as they
would be if fit with Gaussians or Voigt functions); they are skewed blueward in
wavelength, a feature due to the combined variation of ionic abundances with $\xi$
(they decrease faster beyond their maximum value and more gradually prior to that; see
the AMD in {\bf Fig.~1}) and the corresponding variation of the wind velocity with $\xi$, as
expected in a continuous disk wind. The observed skewed  profiles do reflect this
property indicating that the velocity structure of the wind model is a reasonable
representation of the real one. (3) The \fexxv\ feature, usually a strong one in AGNs
\citep{Tombesi13}, is weak in NGC~3783. Our wind model reproduces the weakness of this
line feature, in agreement with the observation.

%[Demos: can you simplify this paragraph...?] \\
In conclusion, the successful modeling of the absorbers of the AGN \obj, along with
those of the Galactic binary GRO~J1655-40 \citep[][]{F17}, with the same global model, argues strongly for an underlying magnetized outflow of an invariant character, not unlike that of our MHD model.
It is found, very similarly to the analysis for GRO~J1655-40, that the wind is magnetically launched and accelerated with  a global density structure of $n(r) = 6.9 \times 10^{11} (r/r_o)^{-1.15}$ cm$^{-3}$ with the viewing angle of $44\deg$. 

We note also that the possibility of MHD-driving disk-winds is independently discussed in the context of the Fe K UFOs in the well-studied Seyfert 1 AGN, NGC~4151, by \cite{KraemerTombesi17}, for example.
Attributed to their typically high ionization parameter (e.g. $\log \xi \gsim 4.0$) and near-Compton thick column (e.g. $N_H \sim 10^{23-24}$ cm$^{-2}$), magnetic-origin seems to be a natural process especially for the UFOs as their study suggests. If a global magnetic field anchored to the underlying accretion disk is a generic component in AGNs/XRBs, then it is quite conceivable that the same magnetic field can play a significant role in launching the WAs discussed in this work for \obj.

%In conclusion, the successful modeling of the absorbers of \obj, along with
%those of GRO J1655-40 \citep{F17}, an object with much different properties from \obj\
%with the same global model while for their different SEDs, argues strongly for an
%underlying outflow of an invariant character, not unlike that of our model.

More detailed spectroscopic analyses of this kind will be made possible with the launch of
{\it XARM} in the coming years and later on by ESA's mission, {\it Athena}, through the micro-calorimeter
observations. These missions will  be able to better constrain the otherwise very
enigmatic absorption properties with an unprecedented statistical significance perhaps
leading to  answering the ultimate question (partially if not fully) of launching
mechanisms and the relations of the UV/X-ray absorbers seen in diverse AGNs.

\acknowledgments 
We thank the anonymous referee for his/her constructive comments and questions. K.F. is grateful to Steve Kraemer, Travis Fischer and Patrick Hall for
their insightful  discussion on magnetically-launched X-ray winds in AGNs. F.T.
acknowledges support by the Programma per Giovani Ricercatori - anno 2014 ``Rita Levi
Montalcini".
The research at the Technion is supported by the I-CORE program of the Planning and Budgeting Committee (grant number 1937/12). EB is grateful for the warm hospitality, support, and visiting professorship at the University of Maryland, College Park, and for funding from the European Union's Horizon 2020 research and innovation programme under the Marie Sklodowska-Curie grant agreement no. 655324.
This work is supported in part by NASA/ADAP (NNH15ZDA001N-ADAP) and {\it
Chandra} AO17 archival proposal grants.

%=============
%
%[Goncalves06; A new model for the Warm Absorber in NGC 3783:
%a single medium in total pressure equilibrium, not a multiple-zone, constant density absorbers]
%
%
%[The thermal instability of the warm absorber in NGC 3783
%R. W. Goosmann1, T. Holczer2, M. Mouchet3, A.-M. Dumont3, E. Behar2, O. Godet4, A. C. Gonalves1, and S. Kaspi2]
%
%
%[AMD by clumpy clouds for Mrk509]
%Adhikari, T. P.; R—?a?ska, A.; Sobolewska, M.; Czerny, B. 2015
%
%============

\end{document}